\documentclass[reprint,nofootinbib,amsmath,amssymb,aps,prl]{revtex4-2}
\usepackage{graphicx}
\usepackage{dcolumn}
\usepackage{bm}

\usepackage{xcolor}
\usepackage{textgreek}
\usepackage[utf8]{inputenc}
\usepackage[english]{babel}
\usepackage{hyperref}
\hypersetup{
    colorlinks=true,
    linkcolor=blue,
    filecolor=blue,
    urlcolor=blue,
    citecolor=blue,
}
\usepackage{color,colortbl}
\usepackage{tensind}
\tensordelimiter{?}
\DeclareGraphicsExtensions{.bmp,.png,.jpg,.pdf}
\usepackage{verbatim}
\usepackage[normalem]{ulem}
\usepackage{orcidlink}
\usepackage{soul}
\usepackage{amsmath}
\usepackage{bm}

\newcommand{\R}{\mathcal{R}}
\newcommand{\D}{\mathcal{D}}
\newcommand{\F}{\mathcal{F}}

\newcommand{\LC}{\mathrm{LC}}
\newcommand{\rlx}{\mathrm{rlx}}

\defcitealias{cohnStellarDistributionBlack1978}{C\&K}

\begin{document}

\preprint{PRL}

\title{Breaking boundaries: extending the orbit-averaged Fokker-Planck equation inside the loss cone}

\author{Luca Broggi\orcidlink{0000-0002-9076-1094}}
\email{luca.broggi@unimib.it}
\affiliation{Dipartimento di Fisica G. Occhialini,
    Università degli Studi di Milano-Bicocca, piazza della Scienza 3, Milano, Italy.}
\affiliation{INFN, Sezione di Milano-Bicocca, Piazza della Scienza 3, I-20126 Milano, Italy}

\begin{abstract}
    In this Letter, we present a new formulation of loss cone theory as a reaction-diffusion system, which accounts for loss cone events through a sink term and can be orbit-averaged. It can recover the standard approach based on boundary conditions, and is derived from a simple physical model that overcomes many of the classical theoretical constraints. We test our formulation by computing the relaxed distribution of disruptive orbits in phase space, that has a simple analytic form and agrees with the pericentre of tidal disruption events at disruption predicted by non-averaged models. This formulation of the problem is particularly suitable for including more physics in tidal disruptions and the analogous problem of gravitational captures, e.g. strong scatterings, gravitational waves emission, physical stellar collisions, and repeating partial disruptions -- that can all act on timescale shorter than two-body relaxation and might cause the tension between the observed vs theoretically predicted population of tidal disruptions.
\end{abstract}

\begin{keywords}
    {galaxies: nuclei, galaxies: star clusters, stellar dynamics, black hole physics.}
\end{keywords}

\maketitle

\paragraph{Introduction.}
\label{sec:intro}
The gravitational field of a massive black hole (MBH) can destroy stars passing too close to it, in a Tidal Disruption Event \citep[TDE;][]{hillsPossiblePowerSource1975,reesTidalDisruptionStars1988,evansTidalDisruptionStar1989,phinneyManifestationsMassiveBlack1989}. These events produce electromagnetic flares so bright that can be observed at cosmological distances \citep{reesTidalDisruptionStars1988, lodatoStellarDisruptionSupermassive2009,strubbeOpticalFlaresTidal2009,rossiProcessStellarTidal2021}, and so numerous that they will probe the dynamical structure of the stellar environment where they are formed \citep{stoneRatesStellarTidal2020,weversMultimessengerAstronomyBlack2023}.

The number of detections of TDEs is growing fast \citep{stoneRatesStellarTidal2020, gezariTidalDisruptionEvents2021}, and is anticipated to increase by orders of magnitude in the coming years thanks to specifically designed observatories such as the Vera Rubin Observatory \citep{thorpTidalDisruptionEvents2019,bricmanProspectsObservingTidal2020} or ULTRASAT \citep{sagivSCIENCEWIDEFIELDUV2014,shvartzvaldULTRASATWidefieldTimedomain2024}. The number of TDEs observed is at the lower-end of theoretical predictions \citep{vanvelzenSeventeenTidalDisruption2021,sazonovFirstTidalDisruption2021,linLuminosityFunctionTidal2022,yaoTidalDisruptionEvent2023}, and a significant fraction of optical and X-Rays TDEs are observed in E+A galaxies, a set of rare post-starburst galaxies showing characteristic absorption lines \citep{arcaviCONTINUUMHeRICHTIDAL2014,frenchTidalDisruptionEvents2016,law-smithTidalDisruptionEvent2017}. Theoretical predictions of the rates are based on relaxation of spherical stellar systems, that continuously pushes stars on deadly orbits around the central MBH  \citep{wangRevisedRatesStellar2004,stoneRatesStellarTidal2016,stoneRatesStellarTidal2020}; recent works suggest that the predicted rate is lower because of dynamical processes limiting it \citep{teboulLossConeShielding2022,broggiRepeatingPartialDisruptions2024}, and that more complex, or time-dependent models of relaxation can explain the E+A preference \citep{stoneDelayTimeDistribution2018,bortolasTidalDisruptionEvents2022,wangExplanationOverrepresentationTidal2024}. The Fokker-Planck formalism employed to compute TDE rates cannot capture all of these process at the same time, and its complicated structure prevents any non-trivial extension. As the same process of relaxation is expected to push compact objects onto orbits plunging on the central MBH \citep{babakScienceSpacebasedInterferometer2017,amaro-seoaneRelativisticDynamicsExtreme2018a,broggiExtremeMassRatio2022,romDynamicsSupermassiveBlack2024}, a satisfactory theoretical base for the computation of TDE rates will naturally correspond to reliable rates of gravitational captures -- the target of future gravitational waves detectors \citep[see e.g.][]{colpiLISADefinitionStudy2024}.

In this Letter, we present an extension of the equation describing the relaxation of spherical stellar systems: the orbit-averaged Fokker-Planck equation (OAFPE). We review the concept of instantaneous disruption of orbits penetrating the disruption radius, also known as the loss cone radius. They define a locus of phase space known as the \textit{loss cone}. Then, we introduce the orbit-average assumption, and explore its impact on loss cone orbits. The paper is organised as follows. First, we introduce the Boltzmann equation (BE) and, assuming spherical symmetry, how it reduces to the OAFPE. Then, we derive our main theoretical result: a sink term to account for the loss cone in the BE, that replaces the need to rely on boundary conditions (BCs) for the orbit-averaged problem. In the standard approximation of nearly radial orbits, the steady-state OAFPE has a simple analytic solution inside the loss cone, and we compare it with the analogous solution of the local equation. Finally, we discuss the impact of this work and its possible applications in the context of the dynamics of spherical dense nuclei.

\paragraph{The OAFPE.}
\label{sec:oaFP}
An isolated stellar system composed of $N$ identical stars around an MBH can be fully described through its distribution function (DF) $f(\bm{x}, \bm{v})$, where $\bm{x}$ and $\bm{v}$ are the position and the velocity that parametrise the phase space of a single star. If the stellar distribution is spherical, the DF has the form \citep{binneyGalacticDynamicsSecond2008}
\begin{equation}\label{eq:DF_def}
    \mathrm{d}N = f(v_r(\bm{x}, \bm{v}), v_t(\bm{x}, \bm{v}), r(\bm{x})) \; \mathrm{d}^3x\,\mathrm{d}^3v
\end{equation}
where $r=\left\lvert \bm{x} \right \rvert$ is the distance from the MBH (the centre of symmetry of the distribution), $v_r = \mathrm{d} r / \mathrm{d} t$ is the radial velocity, and $v_t = \left \lvert \bm{v} - v_r\, \bm{x}/r \right \rvert$ the tangential velocity. In the assumption of orbit-average, one restricts to bound orbits and assumes the form for $f$ prescribed by Jean's theorem
$\mathrm{d}N= f(\mathcal{H}(\bm{x}, \bm{v}), L(\bm{x}, \bm{v}))\; \mathrm{d}^3x\,\mathrm{d}^3v$,
where $\mathcal{H} = \phi(r)+v^2/2$ is the Hamiltonian per unit mass for a star in the system, and $L = r\,v_t$ is the absolute value of the specific orbital angular momentum of a star \citep{binneyGalacticDynamicsSecond2008,merrittDynamicsEvolutionGalactic2013}.

Neglecting collisions, in a time $t$ particles move subject to the potential of the system and $f$ is conserved according to the BE \citep[e.g.][]{binneyGalacticDynamicsSecond2008}. Therefore, it is convenient to express the DF in terms of conserved quantities, e.g. $E = \mathcal{H}$ and $\R = L^2 / L^2_c(E)$, the squared angular momentum scaled to that of the circular orbit. The change of variable is invertible if we consider separately $v_r > 0$ and $v_r < 0$ so that each branch is described by the BE
\begin{equation}
    \frac{\mathrm{d}}{\mathrm{d}t} f = v_r \, \frac{\partial}{\partial r} f + \frac{\partial}{\partial t} f = 0.
\end{equation}
To consider both branches simultaneously, we introduce the time elapsed since the last pericentre $r_p$ (we assume $E<0$)
\begin{equation}\label{eq:tau}
    \tau(r) = \begin{cases}
        \int_{r_p}^{r} \frac{\mathrm{d} r}{v_r} &\quad v_r \geq 0\\
        P + \int_{r_p}^{r} \frac{\mathrm{d}r}{v_r}  &\quad v_r < 0
    \end{cases}
\end{equation}
where $P$ is the orbital radial period. Inserting the collisional effect of weak two-body scatterings, the BE becomes
\begin{equation}\label{eq:local}
    \frac{\partial}{\partial \tau} f + \frac{\partial}{\partial t} f = \Gamma_\mathrm{coll}
\end{equation}
where $\Gamma_\mathrm{coll}$ accounts for the change in orbital parameters and is a total divergence formally linear in $f$ and its partial derivatives up to the second order \citep{cohnStellarDistributionBlack1978}, giving a Fokker-Planck equation. The mean field evolution of a set of particles can be described in terms of the differential distribution in $E$ and $\R$, such that
\begin{equation}
    N = \int  \mathrm{d}E \, \mathrm{d}\R \;n(E,\R; t) \,.
\end{equation}
By direct computation \begin{equation}
n = 4\,\pi^2\,L_c^2\,\int_0^{P} \mathrm{d} \tau f(E,\R, \tau; t) \, ,
\end{equation}
so that $n$ can be interpreted as the orbit-integral of $f$. Applying the orbit-integral $4 \pi^2 L_c^2 \int_0^{P} \mathrm{d} \tau$ to both sides of Eq.~\eqref{eq:local} and assuming $\partial_\tau f = 0$ (\textit{i.e.} replacing $f$ with its orbit-average), we get the OAFPE
\begin{equation}
\begin{split}
     \frac{\partial}{\partial t} n &= 4\,\pi^2\,L_c^2 \, \int_0^{P}\mathrm{d} \tau \,\Gamma_\mathrm{coll}\\
     &\equiv - \bm{\nabla} \cdot \bm{\F}[n]
\end{split}
\end{equation}
where we introduced the particle flux \citep{cohnStellarDistributionBlack1978, merrittDynamicsEvolutionGalactic2013}
\begin{equation}
    \bm{\F} =
    \begin{pmatrix}
    \F_E\\\F_\R
    \end{pmatrix} =
    \begin{pmatrix}
         \D_E f \,+\D_{EE} \,\partial_E f + \D_{E\R} \,\partial_\R f\\
      \D_\R f \,+\D_{E\R} \,\partial_E f + \D_{\R\R} \,\partial_\R f
    \end{pmatrix}   \,.
\end{equation}

\paragraph{Disruptions inside the loss cone}\label{sec:model} Disruptive phenomena happen at a distance smaller than a critical radius $r_\LC$, known as the loss cone radius. For non-relativistic TDEs of stars with mass $m_\star$ and radius $R_\star$
\citep{hillsPossiblePowerSource1975}
\begin{equation}
    r_\LC \simeq R_\star \left({M_\mathrm{BH}}/{m_\star}\right)^{1/3} \,.
\end{equation}
We focus on the process of tidal disruptions, but the same equations describe the rate of gravitational captures, that have a critical radius $r_\mathrm{GC} \simeq 8 \,G\,M_\mathrm{BH}/c^2$  \citep[estimated at the apocentre, in the Newtonian limit;][]{merrittDynamicsEvolutionGalactic2013,mancieriHangingCliffEMRI2024}.
In the standard treatment, a star on an orbit with $r_p<r_\LC$ is destroyed instantaneously as it reaches its pericentre.
Consequently, the distribution in $\tau$ of loss cone orbits ($r_p < r_\LC$ or $\R < \R_\LC$) is no more periodic at $\tau=0$ and $\tau=P$: all particles arriving at the pericentre are destroyed, and no one is found right after the pericentre passage. To compute the rate of disrupted particles,
\citet{cohnStellarDistributionBlack1978} (\citetalias{cohnStellarDistributionBlack1978} from now on) solved the OAFPE outside the loss cone, setting some BCs along this curve. The latter relate the value of the DF on the boundary to the rate of disruptions obtained by solving an approximate, steady-state form of Eq.~\eqref{eq:local} at fixed energy for $\R < \R_\LC$; we summarise their approach in the Appendix \citep[see][for recent implementations]{panFormationRateExtreme2021,broggiExtremeMassRatio2022,wangAccretionmodifiedStellarmassBlack2023}.
It is possible to include the instantaneous disruption at pericentre in the BE as
\begin{equation} \label{eq:sink_term_Boltzmann}
    \frac{\mathrm{d}}{\mathrm{d}t}f(\bm{x},\bm{p}) = \Gamma_\mathrm{coll} - \delta\!\left(\frac{\dot{x}}{\ddot{x}}\right) \Theta(\ddot{x})\,\Theta(r_\LC - x) \, f
\end{equation}
where $x=|\bm{x}|$, $\Theta$ is the Heaviside step function, and $\delta$ is the Dirac function. Without symmetry restrictions, the new term selects local minima of $x$ inside the loss cone, in fact: $\delta(\dot{x}/\ddot{x})$ selects local extrema of $x$\footnote{The denominator $\ddot{x}$ ensures that the $\delta$ has the units of one over time.}, $\Theta(\ddot{x})$ selects minima, and $\Theta(r_\LC-x)$ those inside the loss cone. By considering a spherical system
\begin{equation}\label{eq:local_sink}
    \frac{\partial}{\partial \tau} f + \frac{\partial}{\partial t} f = \Gamma_\mathrm{coll} - \delta(\tau) \Theta(\R_\LC-\R) f\ .
\end{equation}
and assuming $f$ is smooth for $\tau \in [0, P)$, the $\delta$ function requires modifying the boundary condition\footnote{For $r_p>r_\LC$ we recover periodic boundary conditions. For $r_p<r_\LC$ we assume that $\Gamma_\mathrm{coll}$ and $f$ are smooth at $\tau=0$. By integrating the equation in $\tau$ from $0$ to $\epsilon\to0$ we get
$f(\tau=\epsilon) = \epsilon\,(\Gamma_\mathrm{coll} - \partial_tf) + \mathcal{O}(\epsilon^2)$, giving $f(\tau= 0)=0$.
} for $\tau$ as $f(\tau=0)=f(\tau\to P^-)\,\Theta(\R_\LC-\R)$. These are the boundary conditions for the local equation used by \citetalias{cohnStellarDistributionBlack1978}. The rate of destroyed stars is computed by integrating over the full phase space, giving the familiar expression
\begin{equation}\label{eq:FLC_SS}
\begin{split}
    \frac{\mathrm{d} N}{\mathrm{d}t} = - \int_0^{\infty}\mathrm{d}E\int_0^{\R_\LC}\mathrm{d}\R\, 4\pi^2L_c^2 \,f(E,\R,P^-;t) .
\end{split}
\end{equation}

\paragraph{The OAFPE with the loss cone.}

If we now assume $\partial_\tau f=0$ and apply the orbit-integral to Eq.~\eqref{eq:local_sink}, we obtain the \textbf{OAFPE with the loss cone}
\begin{equation}\label{eq:2D_eq_sinkterm}
    \frac{\partial}{\partial t} n = - \bm{\nabla} \cdot \bm{\F}[n] - \frac{ \Theta(\R_\LC-\R)}{P} \, n \,.
\end{equation}
The number of stars on a loss cone orbit destroyed in a time $\delta t$ is
\begin{equation}\label{eq:sink_term}
\begin{split}
    \delta n(E,\R,t) = - n(E,\R; t)\frac{\delta t}{P} \qquad \R<\R_\LC\, .
\end{split}
\end{equation}
In the Appendix we apply this expression in the context of Monte Carlo approaches to the OAFPE \citep{sariTidalDisruptionEvents2019, qunbarEnhancedExtremeMass2023} to account for the loss cone when relaxation changes angular momentum on timescales shorter than $P$. This expression depends only on $E$ and $\R$, and appears as an additional \textit{sink term} in the OAFPE. The BCs for an isolated system are: $\F_E=0$ at $E=0$ and $E\to-\infty$, and $\F_\R=0$ at $\R = 0$ and $\R = 1$. Adding the initial conditions, relaxation in the presence of the loss cone is formulated as a \textit{reaction-diffusion system}, a problem often appearing in the context of chemical and biological models \citep[e.g.][]{kolomgorovStudyDiffusionEquation1937}. In this form, relaxation and disruptions are included through separate, independent terms; instead, classical loss cone BCs must consider relaxation to model disruptions.

\label{sec:comparison}
\paragraph{Steady-state, nearly radial limit.} To test the new formulation of the problem, we address the simplified problem to find the steady distribution of loss cone orbits, following \citetalias{cohnStellarDistributionBlack1978}. For small $\R$
\begin{equation}\label{eq:FP_inside}
\begin{split}
    -\bm{\nabla}\cdot \bm{\F}\simeq-\partial_\R\F_\R &\simeq 4\,\pi^2\,L_c^2\,\int_0^{P} \mathrm{d} \tau \; d(E,\tau)\,\partial_\R [\R\, \partial_\R f]\\
    &= 4\,\pi^2\,L_c^2\,P\, \D(E) \,\partial_\R [\R\, \partial_\R f]
\end{split}
\end{equation}
where $d(E,\tau)$ is a local diffusion coefficient, and $\D(E)$ its orbit-average.
The boundary conditions for the problem inside the loss cone are $\F_\R(\R=0)=0$ and $f(E,\R_\LC) = f_\LC(E)$.

\begin{figure*}\centering
    \includegraphics[width=0.9\linewidth]{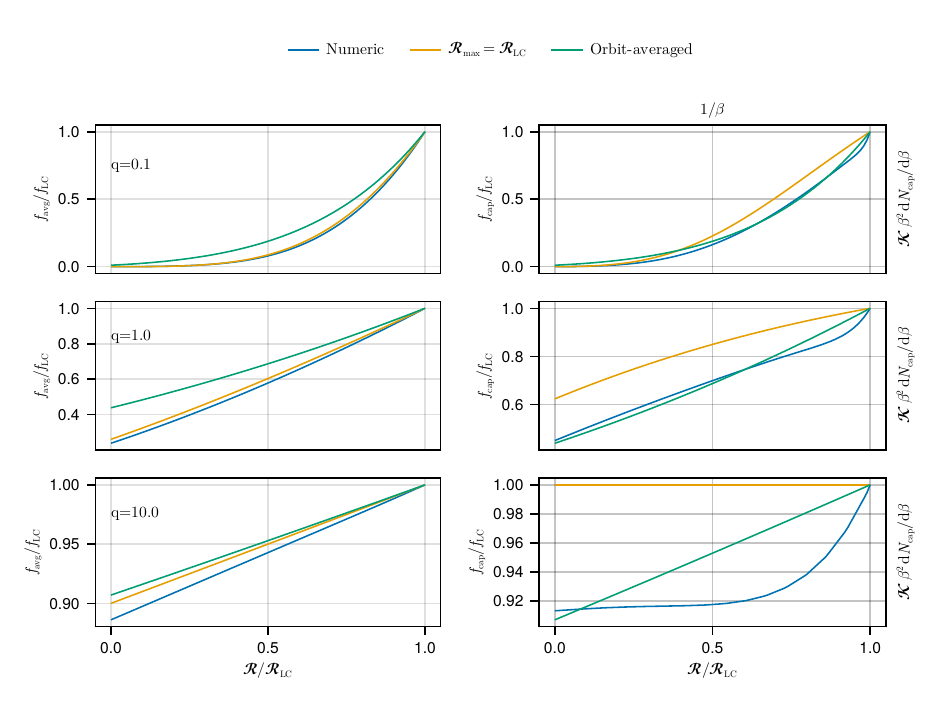}
    \caption{\label{fig:comp_CK_MM} Comparison between the solution of the local Fokker-Planck equation inside the loss cone (\textit{numeric}, blue) with the analytic limit presented in \citet{merrittDynamicsEvolutionGalactic2013} ($\R_\mathrm{max} = \R_\LC$, orange) and our orbit-averaged estimate (\textit{orbit-averaged}, Eq.~\eqref{eq:inside}, green). On the left, we show the average distribution inside the loss cone $f_\mathrm{avg}$, on the right the distribution of $\R/\R_\LC$ at capture $f_\mathrm{cap}$ for stars on orbits with different $q$. For very eccentric Keplerian orbits $\R/\R_\LC \simeq 1/\beta$, and the distribution of penetration factor of TDEs at a given $q$ (Eq.~\ref{eq:peri_distr}) is given by $f_\mathrm{capt} / f_\LC= \mathcal{K}\,\beta^2\, dN_\mathrm{capt}/d\beta$ where $\mathcal{K}$ depends on normalisation. We consider three values of the loss cone diffusivity parameter (Eq.~\ref{eq:q}) $q=0.1$ (first row), $q=1.0$ (second row), and $q=10.0$ (third row). Each curve is normalised to its value at $\R_\LC$.}
\end{figure*}

From Eqs. (\ref{eq:2D_eq_sinkterm}-\ref{eq:FP_inside}) we obtain a partial differential equation for $f(E; \R, t)$. By setting $\partial_t n = 0$, it becomes an ordinary differential equation\footnote{\citet{teboulLossConeShielding2022} follow a similar approach to derive the quasi-relaxed distribution in angular momentum when strong scatterings are dominant -- but the equation has no simple analytic solution. This suggests the possibility of a simple, combined treatment of weak and strong scatterings.}
\begin{equation}\label{eq:ODE}
    \frac{\partial}{\partial \R} \R \frac{\partial}{\partial \R} f = \frac{\Theta(\R_\LC-\R)}{\D\, P}\,f \ .
\end{equation}
Finally, by inserting Eq.~\eqref{eq:ODE} in Eq.~\eqref{eq:FP_inside} and integrating from 0 to $\R_\LC$ we can write
\begin{equation}
    \F_\R^\LC = - 4\,\pi^2\,L_c^2\, \int_0^{\R_\LC}\mathrm{d}\R \, f(E,\R) \,.
\end{equation}

The combination $\D\,P$ appearing in Eq.~\eqref{eq:ODE} is the expected variation $\Delta \R$ over an orbit with $\R=\R_\LC$, and quantifies the strength of scatterings. It is usually parametrised through the adimensional \textit{loss cone diffusivity parameter}
\begin{equation}\label{eq:q}
    q = \frac{\D\, P }{\R_\LC} = \left. \frac{P}{t_\rlx}\right \rvert_{\R_\LC}
\end{equation}
where $t_\rlx = \R / \D$ is the relaxation time of an orbit, i.e. the time required for an orbit to change its angular momentum by the order of itself. When $q \gg 1$, stars will spend on average only a time $t_\rlx = P/q$ on an orbit grazing the loss cone before being scattered away; this is known as the \textit{full} or \textit{pinhole} loss cone regime as few particles reach disruption. When $q \ll 1$, stars will slightly change $\R$ over a single period; most of them will be destroyed as soon as their angular momentum is $\R_\LC$ -- this is the \textit{empty loss cone} or \textit{diffusive} regime.

The solution to Eq.~\eqref{eq:ODE} is:
\begin{align}\label{eq:inside}
    f &= \begin{cases}
        \frac{f_\LC}{I_0(2/\sqrt{q})} \, I_0\left(\frac{2}{\sqrt{q}}\,\sqrt{\frac{\mathcal{R}}{\mathcal{R}_\mathrm{lc}}}\right) \quad &\R<\R_\LC\\
        f_\LC\,\left[ \frac{I_1(2/\sqrt{q})}{\sqrt{q}\,I_0(2/\sqrt{q})} \, \log\frac{\R}{R_\LC} + 1\right] &\R \geq \R_\LC
    \end{cases}\\
    \partial_\mathcal{R} f &= \begin{cases}\, \frac{f_\mathrm{lc}}{\sqrt{q\, \mathcal{R}_\mathrm{lc} }} \,\frac{1}{I_0(2/\sqrt{q})} \, \frac{1}{\sqrt{\R}}\, I_1\left(\frac{2}{\sqrt{q}}\,\sqrt{\frac{\mathcal{R}}{\mathcal{R}_\mathrm{lc}}}\right) \, \quad &\R<\R_\LC\\
    \frac{f_\LC\,I_1(2/\sqrt{q})}{\sqrt{q}\,I_0(2/\sqrt{q})} \frac{1}{\R} &\R \geq \R_\LC
    \end{cases}
\end{align}
where $I_k(z)$ is the modified Bessel function of the first kind of order $k$. The total flux entering the loss cone is
\begin{equation}
\begin{split}\label{eq:Flux}
    \F_\R^\LC= 4\,\pi^2\,L_c^2\,\R_\LC\,f_\LC\,\sqrt{q}\,I_1(2/\sqrt{q}) \, /\, I_0(2/\sqrt{q}) \,.
\end{split}
\end{equation}

Eq.~\eqref{eq:inside} represents both the average DF inside the loss cone and the distribution in $\R$ of stars destroyed with energy $E$ (see Eq.~\ref{eq:FLC_SS}).
The latter is directly related to the orbital parameters of stars in TDEs or, when considering the gravitational capture of compact objects, in plunges. The pericentre distribution of TDEs at energy $E$ must be computed accounting for the stellar potential, as it affects the angular momentum of the circular orbit. Considering very eccentric orbits and neglecting the stellar potential, the pericentre distribution is obtained by substituting $\R = 2\,r_p/r_c(E)$, where $r_c(E)$ is the radius of the circular orbit with energy $E$, in Eq.~\eqref{eq:inside} for $\R<\R_\LC$.
Alternatively, the \textbf{distribution of the penetration factor} $\beta = r_\LC / r_p$, is
\begin{equation}\label{eq:peri_distr}
    \mathrm{d} N_\mathrm{cap} = \tilde{\mathcal{N}} \,I_0\left({2}/{\sqrt{q\,\beta}}\right) \,\mathrm{d}\beta/\beta^2 \qquad 1 \leq \beta < \infty
\end{equation}
where $\tilde{\mathcal{N}}=[\sqrt{q}\, I_1(2/\sqrt{q})]^{-1}$ ensures the normalisation of the distribution to 1 (see the right column of Fig.~\ref{fig:comp_CK_MM}).

\paragraph{Impact of orbit-averaging inside the loss cone in the steady-state limit}
We now compare Eq.~\eqref{eq:inside} with the predictions from the local equation \citep{cohnStellarDistributionBlack1978,merrittDynamicsEvolutionGalactic2013}.
\citetalias{cohnStellarDistributionBlack1978} find the solution $f(\R, \tau)$ by solving numerically the local equation  Eq.~\eqref{eq:local_sink} in the steady-state, nearly radial limit up to\footnote{The maximum value of $\R$ must be large enough that the solution inside the loss cone is insensitive to it, since $f$ is set to be uniform in $\tau$ here. See the Appendix for our estimate $\R_\mathrm{max}$} $\R_\mathrm{max}\simeq (1+5q)\,\R_\LC$. A closed-form approximant is built assuming $\R_\mathrm{max} = \R_\LC$, we report the expression of the average $f^\mathrm{M}_\mathrm{avg}$ in Eq.~\eqref{eq:avg_local_FP} and the distribution at disruption $f^\mathrm{M}_\mathrm{cap}$ in Eq.~\eqref{eq:cap_local_FP} in the Appendix. These approximants have been widely used to estimate the pericentre distribution of stars \citep[e.g.][]{pfisterObservableGravitationalWaves2022,changRatesStellarTidal2024}, but their slow convergence makes them computationally expensive and are often replaced by simpler formulas \citep[e.g.][]{stoneRatesStellarTidal2020,bortolasPartialStellarTidal2023}.

In Fig.~\ref{fig:comp_CK_MM} we compare the \citetalias{cohnStellarDistributionBlack1978} solution of the local Fokker-Planck equation with the orbit-averaged Eq.~\eqref{eq:inside}, and the $\R_\mathrm{max} = \R_\LC$ approximants. All curves are peaked at $\R=\R_\LC$ for $q=0.1$, and are nearly flat for $q=10.0$. The orbit-averaged solution differs most from the numeric solution at $q=1.0$, with the largest deviation of $80\%$; the distribution in $\tau$ plays an important role for $t_\mathrm{rlx}=P$. The $\R_\mathrm{max} = \R_\LC$ approximation performs worst at predicting the capture distribution with $q=1.0$, with a maximum error of $40\%$; here $\R_\mathrm{max} = \R_\LC$ is a poor approximation\footnote{The approximant for $\R=\R_\mathrm{rlx}$ is worse at predicting the $\R$ distribution at disruption than the average distribution. In fact, the numerical solution shows that at any value of $q$ the distribution at $\R_\LC$ is not uniform in $\tau$ (see the Appendix).}.

In the Appendix we compare the loss cone fluxes and find the orbit-averaged solution worse than the $\R_\mathrm{max}=\R_\LC$ approximation at any $q$, with an overestimation of $15-35\%$ and $5-20\%$ respectively. Therefore, the new formalism should be used through the sink term, while for this problem classical BCs at $\R_\LC$ are better.

\paragraph{Discussion and conclusions.}
\label{sec:discussion}
The formation rate of TDEs inferred from observations is lower than theoretical predictions based on two body relaxation and loss cone theory, even assuming spherical symmetry around a single MBH (axisymmetric potentials and central MBH binaries both increase the rate \citep[e.g.][]{vasilievLossconeProblemAxisymmetric2013, mazzolariExtremeMassRatio2022, liuTidalDisruptionEvents2023}). The current approach is based on the orbit-averaged Fokker-Planck equation (OAFPE), and requires the assumption that orbits penetrating the loss cone are nearly radial and subject only to two-body scatterings to compute their steady-state distribution in angular momentum.

We derived the orbit-averaged, time-dependent treatment of loss cone orbits extending the OAFPE inside the loss cone. Starting from the inclusion of a sink term in the Boltzmann equation (Eqs.~\ref{eq:sink_term_Boltzmann},\ref{eq:local_sink},\ref{eq:2D_eq_sinkterm}),  we formulate the problem of relaxation with loss cone disruptions as a \textit{reaction-diffusion} system. Compared to classical boundary conditions, it separates the contribution of disruptions and relaxation effects, enabling simpler modelling of new physics, and removes the hypotheses of instantaneous relaxation and extreme eccentricity of loss cone orbits. In the Appendix, we show a direct application of the orbit-averaged model to Monte Carlo simulations of two-body relaxation.

We tested this formulation in the problem of finding the orbit-averaged steady-state equation of nearly radial orbits, and found that orbit-averaging inside the loss cone performs worst at $q=1$, and other approximants might perform better for two-body relaxation only. However, additional processes might affect the angular momentum of stars and the orbit-averaged approximation might be better suited. In these situations, the time-dependent sink term can account for disruptions, or equations like
Eq.~\eqref{eq:ODE} can extend the classical approach based on BCs. Physical phenomena acting on timescales shorter than $t_\mathrm{rlx}$ include strong scatterings \citep{teboulLossConeShielding2022, kaurExtrememassRatioInspirals2024}, extreme emission of gravitational waves \citep{ qunbarEnhancedExtremeMass2023,kaurSemiAnalyticalFokkerPlanck2024,mancieriHangingCliffEMRI2024}, physical stellar collisions \citep{sariTidalDisruptionEvents2019}, and partial disruptions \citep{broggiRepeatingPartialDisruptions2024}. Some of these might explain the population of TDEs currently observed, offering the possibility of testing in time a more complex and realistic model for the relaxation of galactic nuclei, with a consequent more in-depth understanding of the process.

In the time-dependent problem, the sink term can be included in the ergodic Fokker-Planck equation. This is a 1D evolution equation for ergodic DFs mostly used to study quasi-relaxed profiles \citep{bahcallStarDistributionMassive1977,linialStellarDistributionsSupermassive2022,romEnergyFluxParticle2023}, which can estimate time-dependent loss cone rates assuming instantaneous relaxation in angular momentum \citep[e.g. in \textsc{PhaseFlow}:][]{vasilievNewFokkerPlanckApproach2017}. The sink term can account for the depletion of orbits with semimajor axis comparable to or smaller than $r_\LC$, as happens when stellar components have different loss cone radii (e.g. stellar binaries and stars). Depending on $r_\LC$, the sink term $\propto -1/P$ will deplete stars inside the loss cone on the expected timescale.

Eq.~\eqref{eq:2D_eq_sinkterm} is derived for a population of identical stars. However, it is easily extended to an arbitrary number of stellar components: each of them will evolve according to a formally identical equation, and the argument of the $\Theta$ will change based on the loss cone process of that component.

We remark that the diffusion of loss cone orbits depends on time and on the radial phase, and cannot be captured exactly with either the steady-state ($\partial_t f$=0) or the orbit-average ($\partial_\tau f$=0) assumption. The application of the OAFPE to realistic galaxies is not immediate, because of the possibility of radial-dependent effects induced by the environment or by axi-symmetry (e.g. chaotic behaviour in the galactic outskirts \cite{holley-bockelmannFullLossCone2006,lezhninTIDALDISRUPTIONRATES2016} and the presence of gas and disk-like structures in the inner region \cite{panFormationRateExtreme2021,roznerStellarDistributionsSupermassive2025}), even though the rate of TDEs might be weakly affected for $M_\bullet \lesssim 10^7 M_\odot$ \citep{vasilievLossconeProblemAxisymmetric2013, vasilievRatesCaptureStars2014}.



\begin{acknowledgments}
The author acknowledges enlightening conversations with A. Sesana, N.C. Stone, M. Bonetti, M. Dotti, and the entire B Massive group; and the early enthusiasm of E.M. Rossi, O. Teboul, K. Kaur and P. Martire when discussing the idea of this work in Crete. The author thanks the anonymous referees for their comments that helped to improve the quality of this work.
The author acknowledges the financial support provided under the European Union’s H2020 ERC Consolidator Grant “Binary Massive Black Hole Astrophysics” (B Massive, Grant Agreement: 818691).\\
\textit{Software}: the Julia language \citep{bezansonJuliaFreshApproach2017}; Makie.jl \citep{danischMakieJlFlexible2021}; Interpolations.jl \cite{kittisopikulJuliaMathInterpolationsJl2023}.
\end{acknowledgments}

\bibliography{apssamp}

\begin{thebibliography}{63}%
\makeatletter
\providecommand \@ifxundefined [1]{%
 \@ifx{#1\undefined}
}%
\providecommand \@ifnum [1]{%
 \ifnum #1\expandafter \@firstoftwo
 \else \expandafter \@secondoftwo
 \fi
}%
\providecommand \@ifx [1]{%
 \ifx #1\expandafter \@firstoftwo
 \else \expandafter \@secondoftwo
 \fi
}%
\providecommand \natexlab [1]{#1}%
\providecommand \enquote  [1]{``#1''}%
\providecommand \bibnamefont  [1]{#1}%
\providecommand \bibfnamefont [1]{#1}%
\providecommand \citenamefont [1]{#1}%
\providecommand \href@noop [0]{\@secondoftwo}%
\providecommand \href [0]{\begingroup \@sanitize@url \@href}%
\providecommand \@href[1]{\@@startlink{#1}\@@href}%
\providecommand \@@href[1]{\endgroup#1\@@endlink}%
\providecommand \@sanitize@url [0]{\catcode `\\12\catcode `\$12\catcode `\&12\catcode `\#12\catcode `\^12\catcode `\_12\catcode `\%12\relax}%
\providecommand \@@startlink[1]{}%
\providecommand \@@endlink[0]{}%
\providecommand \url  [0]{\begingroup\@sanitize@url \@url }%
\providecommand \@url [1]{\endgroup\@href {#1}{\urlprefix }}%
\providecommand \urlprefix  [0]{URL }%
\providecommand \Eprint [0]{\href }%
\providecommand \doibase [0]{http://dx.doi.org/}%
\providecommand \selectlanguage [0]{\@gobble}%
\providecommand \bibinfo  [0]{\@secondoftwo}%
\providecommand \bibfield  [0]{\@secondoftwo}%
\providecommand \translation [1]{[#1]}%
\providecommand \BibitemOpen [0]{}%
\providecommand \bibitemStop [0]{}%
\providecommand \bibitemNoStop [0]{.\EOS\space}%
\providecommand \EOS [0]{\spacefactor3000\relax}%
\providecommand \BibitemShut  [1]{\csname bibitem#1\endcsname}%
\let\auto@bib@innerbib\@empty
\bibitem [{\citenamefont {Hills}(1975)}]{hillsPossiblePowerSource1975}%
  \BibitemOpen
  \bibfield  {author} {\bibinfo {author} {\bibfnamefont {J.~G.}\ \bibnamefont {Hills}},\ }\href {\doibase 10.1038/254295a0} {\bibfield  {journal} {\bibinfo  {journal} {Nature}\ }\textbf {\bibinfo {volume} {254}},\ \bibinfo {pages} {295} (\bibinfo {year} {1975})}\BibitemShut {NoStop}%
\bibitem [{\citenamefont {Rees}(1988)}]{reesTidalDisruptionStars1988}%
  \BibitemOpen
  \bibfield  {author} {\bibinfo {author} {\bibfnamefont {M.~J.}\ \bibnamefont {Rees}},\ }\href {\doibase 10.1038/333523a0} {\bibfield  {journal} {\bibinfo  {journal} {Nature}\ }\textbf {\bibinfo {volume} {333}},\ \bibinfo {pages} {523} (\bibinfo {year} {1988})}\BibitemShut {NoStop}%
\bibitem [{\citenamefont {Evans}\ and\ \citenamefont {Kochanek}(1989)}]{evansTidalDisruptionStar1989}%
  \BibitemOpen
  \bibfield  {author} {\bibinfo {author} {\bibfnamefont {C.~R.}\ \bibnamefont {Evans}}\ and\ \bibinfo {author} {\bibfnamefont {C.~S.}\ \bibnamefont {Kochanek}},\ }\href {\doibase 10.1086/185567} {\bibfield  {journal} {\bibinfo  {journal} {The Astrophysical Journal}\ }\textbf {\bibinfo {volume} {346}},\ \bibinfo {pages} {L13} (\bibinfo {year} {1989})}\BibitemShut {NoStop}%
\bibitem [{\citenamefont {Phinney}(1989)}]{phinneyManifestationsMassiveBlack1989}%
  \BibitemOpen
  \bibfield  {author} {\bibinfo {author} {\bibfnamefont {E.~S.}\ \bibnamefont {Phinney}},\ }\href@noop {} {\ \textbf {\bibinfo {volume} {136}},\ \bibinfo {pages} {543} (\bibinfo {year} {1989})}\BibitemShut {NoStop}%
\bibitem [{\citenamefont {Lodato}\ \emph {et~al.}(2009)\citenamefont {Lodato}, \citenamefont {King},\ and\ \citenamefont {Pringle}}]{lodatoStellarDisruptionSupermassive2009}%
  \BibitemOpen
  \bibfield  {author} {\bibinfo {author} {\bibfnamefont {G.}~\bibnamefont {Lodato}}, \bibinfo {author} {\bibfnamefont {A.~R.}\ \bibnamefont {King}}, \ and\ \bibinfo {author} {\bibfnamefont {J.~E.}\ \bibnamefont {Pringle}},\ }\href {\doibase 10.1111/j.1365-2966.2008.14049.x} {\bibfield  {journal} {\bibinfo  {journal} {Monthly Notices of the Royal Astronomical Society}\ }\textbf {\bibinfo {volume} {392}},\ \bibinfo {pages} {332} (\bibinfo {year} {2009})}\BibitemShut {NoStop}%
\bibitem [{\citenamefont {Strubbe}\ and\ \citenamefont {Quataert}(2009)}]{strubbeOpticalFlaresTidal2009}%
  \BibitemOpen
  \bibfield  {author} {\bibinfo {author} {\bibfnamefont {L.~E.}\ \bibnamefont {Strubbe}}\ and\ \bibinfo {author} {\bibfnamefont {E.}~\bibnamefont {Quataert}},\ }\href {\doibase 10.1111/j.1365-2966.2009.15599.x} {\bibfield  {journal} {\bibinfo  {journal} {Monthly Notices of the Royal Astronomical Society}\ }\textbf {\bibinfo {volume} {400}},\ \bibinfo {pages} {2070} (\bibinfo {year} {2009})}\BibitemShut {NoStop}%
\bibitem [{\citenamefont {Rossi}\ \emph {et~al.}(2021)\citenamefont {Rossi}, \citenamefont {Stone}, \citenamefont {{Law-Smith}}, \citenamefont {Macleod}, \citenamefont {Lodato}, \citenamefont {Dai},\ and\ \citenamefont {Mandel}}]{rossiProcessStellarTidal2021}%
  \BibitemOpen
  \bibfield  {author} {\bibinfo {author} {\bibfnamefont {E.~M.}\ \bibnamefont {Rossi}}, \bibinfo {author} {\bibfnamefont {N.~C.}\ \bibnamefont {Stone}}, \bibinfo {author} {\bibfnamefont {J.~A.~P.}\ \bibnamefont {{Law-Smith}}}, \bibinfo {author} {\bibfnamefont {M.}~\bibnamefont {Macleod}}, \bibinfo {author} {\bibfnamefont {G.}~\bibnamefont {Lodato}}, \bibinfo {author} {\bibfnamefont {J.~L.}\ \bibnamefont {Dai}}, \ and\ \bibinfo {author} {\bibfnamefont {I.}~\bibnamefont {Mandel}},\ }\href {\doibase 10.1007/s11214-021-00818-7} {\bibfield  {journal} {\bibinfo  {journal} {Space Science Reviews}\ }\textbf {\bibinfo {volume} {217}},\ \bibinfo {pages} {40} (\bibinfo {year} {2021})}\BibitemShut {NoStop}%
\bibitem [{\citenamefont {Stone}\ \emph {et~al.}(2020)\citenamefont {Stone}, \citenamefont {Vasiliev}, \citenamefont {Kesden}, \citenamefont {Rossi}, \citenamefont {Perets},\ and\ \citenamefont {{Amaro-Seoane}}}]{stoneRatesStellarTidal2020}%
  \BibitemOpen
  \bibfield  {author} {\bibinfo {author} {\bibfnamefont {N.~C.}\ \bibnamefont {Stone}}, \bibinfo {author} {\bibfnamefont {E.}~\bibnamefont {Vasiliev}}, \bibinfo {author} {\bibfnamefont {M.}~\bibnamefont {Kesden}}, \bibinfo {author} {\bibfnamefont {E.~M.}\ \bibnamefont {Rossi}}, \bibinfo {author} {\bibfnamefont {H.~B.}\ \bibnamefont {Perets}}, \ and\ \bibinfo {author} {\bibfnamefont {P.}~\bibnamefont {{Amaro-Seoane}}},\ }\href {\doibase 10.1007/s11214-020-00651-4} {\bibfield  {journal} {\bibinfo  {journal} {Space Science Reviews}\ }\textbf {\bibinfo {volume} {216}},\ \bibinfo {pages} {35} (\bibinfo {year} {2020})},\ \Eprint {http://arxiv.org/abs/2003.08953} {arXiv:2003.08953} \BibitemShut {NoStop}%
\bibitem [{\citenamefont {Wevers}\ and\ \citenamefont {Ryu}(2023)}]{weversMultimessengerAstronomyBlack2023}%
  \BibitemOpen
  \bibfield  {author} {\bibinfo {author} {\bibfnamefont {T.}~\bibnamefont {Wevers}}\ and\ \bibinfo {author} {\bibfnamefont {T.}~\bibnamefont {Ryu}},\ }\href {\doibase 10.48550/arXiv.2310.16879} {\enquote {\bibinfo {title} {Multi-messenger astronomy with black holes: Tidal disruption events},}\ } (\bibinfo {year} {2023})\BibitemShut {NoStop}%
\bibitem [{\citenamefont {Gezari}(2021)}]{gezariTidalDisruptionEvents2021}%
  \BibitemOpen
  \bibfield  {author} {\bibinfo {author} {\bibfnamefont {S.}~\bibnamefont {Gezari}},\ }\href {\doibase 10.1146/annurev-astro-111720-030029} {\bibfield  {journal} {\bibinfo  {journal} {Annual Review of Astronomy and Astrophysics}\ }\textbf {\bibinfo {volume} {59}},\ \bibinfo {pages} {21} (\bibinfo {year} {2021})}\BibitemShut {NoStop}%
\bibitem [{\citenamefont {Thorp}\ \emph {et~al.}(2019)\citenamefont {Thorp}, \citenamefont {Chadwick},\ and\ \citenamefont {Sesana}}]{thorpTidalDisruptionEvents2019}%
  \BibitemOpen
  \bibfield  {author} {\bibinfo {author} {\bibfnamefont {S.}~\bibnamefont {Thorp}}, \bibinfo {author} {\bibfnamefont {E.}~\bibnamefont {Chadwick}}, \ and\ \bibinfo {author} {\bibfnamefont {A.}~\bibnamefont {Sesana}},\ }\href {\doibase 10.1093/mnras/stz1970} {\bibfield  {journal} {\bibinfo  {journal} {Monthly Notices of the Royal Astronomical Society}\ }\textbf {\bibinfo {volume} {488}},\ \bibinfo {pages} {4042} (\bibinfo {year} {2019})}\BibitemShut {NoStop}%
\bibitem [{\citenamefont {Bricman}\ and\ \citenamefont {Gomboc}(2020)}]{bricmanProspectsObservingTidal2020}%
  \BibitemOpen
  \bibfield  {author} {\bibinfo {author} {\bibfnamefont {K.}~\bibnamefont {Bricman}}\ and\ \bibinfo {author} {\bibfnamefont {A.}~\bibnamefont {Gomboc}},\ }\href {\doibase 10.3847/1538-4357/ab6989} {\bibfield  {journal} {\bibinfo  {journal} {The Astrophysical Journal}\ }\textbf {\bibinfo {volume} {890}},\ \bibinfo {pages} {73} (\bibinfo {year} {2020})}\BibitemShut {NoStop}%
\bibitem [{\citenamefont {Sagiv}\ \emph {et~al.}(2014)\citenamefont {Sagiv}, \citenamefont {{Gal-Yam}}, \citenamefont {Ofek}, \citenamefont {Waxman}, \citenamefont {Aharonson}, \citenamefont {Kulkarni}, \citenamefont {Nakar}, \citenamefont {Maoz}, \citenamefont {Trakhtenbrot}, \citenamefont {Phinney}, \citenamefont {Topaz}, \citenamefont {Beichman}, \citenamefont {Murthy},\ and\ \citenamefont {Worden}}]{sagivSCIENCEWIDEFIELDUV2014}%
  \BibitemOpen
  \bibfield  {author} {\bibinfo {author} {\bibfnamefont {I.}~\bibnamefont {Sagiv}}, \bibinfo {author} {\bibfnamefont {A.}~\bibnamefont {{Gal-Yam}}}, \bibinfo {author} {\bibfnamefont {E.~O.}\ \bibnamefont {Ofek}}, \bibinfo {author} {\bibfnamefont {E.}~\bibnamefont {Waxman}}, \bibinfo {author} {\bibfnamefont {O.}~\bibnamefont {Aharonson}}, \bibinfo {author} {\bibfnamefont {S.~R.}\ \bibnamefont {Kulkarni}}, \bibinfo {author} {\bibfnamefont {E.}~\bibnamefont {Nakar}}, \bibinfo {author} {\bibfnamefont {D.}~\bibnamefont {Maoz}}, \bibinfo {author} {\bibfnamefont {B.}~\bibnamefont {Trakhtenbrot}}, \bibinfo {author} {\bibfnamefont {E.~S.}\ \bibnamefont {Phinney}}, \bibinfo {author} {\bibfnamefont {J.}~\bibnamefont {Topaz}}, \bibinfo {author} {\bibfnamefont {C.}~\bibnamefont {Beichman}}, \bibinfo {author} {\bibfnamefont {J.}~\bibnamefont {Murthy}}, \ and\ \bibinfo {author} {\bibfnamefont {S.~P.}\ \bibnamefont {Worden}},\ }\href {\doibase 10.1088/0004-6256/147/4/79} {\bibfield  {journal} {\bibinfo  {journal} {The
  Astronomical Journal}\ }\textbf {\bibinfo {volume} {147}},\ \bibinfo {pages} {79} (\bibinfo {year} {2014})}\BibitemShut {NoStop}%
\bibitem [{\citenamefont {Shvartzvald}\ \emph {et~al.}(2024)\citenamefont {Shvartzvald}, \citenamefont {Waxman}, \citenamefont {{Gal-Yam}}, \citenamefont {Ofek}, \citenamefont {{Ben-Ami}}, \citenamefont {Berge}, \citenamefont {Kowalski}, \citenamefont {B{\"u}hler}, \citenamefont {Worm}, \citenamefont {Rhoads}, \citenamefont {Arcavi}, \citenamefont {Maoz}, \citenamefont {Polishook}, \citenamefont {Stone}, \citenamefont {Trakhtenbrot}, \citenamefont {Ackermann}, \citenamefont {Aharonson}, \citenamefont {Birnholtz}, \citenamefont {Chelouche}, \citenamefont {Guetta}, \citenamefont {Hallakoun}, \citenamefont {Horesh}, \citenamefont {Kushnir}, \citenamefont {Mazeh}, \citenamefont {Nordin}, \citenamefont {Ofir}, \citenamefont {Ohm}, \citenamefont {Parsons}, \citenamefont {Pe'er}, \citenamefont {Perets}, \citenamefont {Perdelwitz}, \citenamefont {Poznanski}, \citenamefont {Sadeh}, \citenamefont {Sagiv}, \citenamefont {Shahaf}, \citenamefont {Soumagnac}, \citenamefont {{Tal-Or}}, \citenamefont {Santen}, \citenamefont
  {Zackay}, \citenamefont {Guttman}, \citenamefont {Rekhi}, \citenamefont {Townsend}, \citenamefont {Weinstein},\ and\ \citenamefont {Wold}}]{shvartzvaldULTRASATWidefieldTimedomain2024}%
  \BibitemOpen
  \bibfield  {author} {\bibinfo {author} {\bibfnamefont {Y.}~\bibnamefont {Shvartzvald}}, \bibinfo {author} {\bibfnamefont {E.}~\bibnamefont {Waxman}}, \bibinfo {author} {\bibfnamefont {A.}~\bibnamefont {{Gal-Yam}}}, \bibinfo {author} {\bibfnamefont {E.~O.}\ \bibnamefont {Ofek}}, \bibinfo {author} {\bibfnamefont {S.}~\bibnamefont {{Ben-Ami}}}, \bibinfo {author} {\bibfnamefont {D.}~\bibnamefont {Berge}}, \bibinfo {author} {\bibfnamefont {M.}~\bibnamefont {Kowalski}}, \bibinfo {author} {\bibfnamefont {R.}~\bibnamefont {B{\"u}hler}}, \bibinfo {author} {\bibfnamefont {S.}~\bibnamefont {Worm}}, \bibinfo {author} {\bibfnamefont {J.~E.}\ \bibnamefont {Rhoads}}, \bibinfo {author} {\bibfnamefont {I.}~\bibnamefont {Arcavi}}, \bibinfo {author} {\bibfnamefont {D.}~\bibnamefont {Maoz}}, \bibinfo {author} {\bibfnamefont {D.}~\bibnamefont {Polishook}}, \bibinfo {author} {\bibfnamefont {N.}~\bibnamefont {Stone}}, \bibinfo {author} {\bibfnamefont {B.}~\bibnamefont {Trakhtenbrot}}, \bibinfo {author} {\bibfnamefont
  {M.}~\bibnamefont {Ackermann}}, \bibinfo {author} {\bibfnamefont {O.}~\bibnamefont {Aharonson}}, \bibinfo {author} {\bibfnamefont {O.}~\bibnamefont {Birnholtz}}, \bibinfo {author} {\bibfnamefont {D.}~\bibnamefont {Chelouche}}, \bibinfo {author} {\bibfnamefont {D.}~\bibnamefont {Guetta}}, \bibinfo {author} {\bibfnamefont {N.}~\bibnamefont {Hallakoun}}, \bibinfo {author} {\bibfnamefont {A.}~\bibnamefont {Horesh}}, \bibinfo {author} {\bibfnamefont {D.}~\bibnamefont {Kushnir}}, \bibinfo {author} {\bibfnamefont {T.}~\bibnamefont {Mazeh}}, \bibinfo {author} {\bibfnamefont {J.}~\bibnamefont {Nordin}}, \bibinfo {author} {\bibfnamefont {A.}~\bibnamefont {Ofir}}, \bibinfo {author} {\bibfnamefont {S.}~\bibnamefont {Ohm}}, \bibinfo {author} {\bibfnamefont {D.}~\bibnamefont {Parsons}}, \bibinfo {author} {\bibfnamefont {A.}~\bibnamefont {Pe'er}}, \bibinfo {author} {\bibfnamefont {H.~B.}\ \bibnamefont {Perets}}, \bibinfo {author} {\bibfnamefont {V.}~\bibnamefont {Perdelwitz}}, \bibinfo {author} {\bibfnamefont
  {D.}~\bibnamefont {Poznanski}}, \bibinfo {author} {\bibfnamefont {I.}~\bibnamefont {Sadeh}}, \bibinfo {author} {\bibfnamefont {I.}~\bibnamefont {Sagiv}}, \bibinfo {author} {\bibfnamefont {S.}~\bibnamefont {Shahaf}}, \bibinfo {author} {\bibfnamefont {M.}~\bibnamefont {Soumagnac}}, \bibinfo {author} {\bibfnamefont {L.}~\bibnamefont {{Tal-Or}}}, \bibinfo {author} {\bibfnamefont {J.~V.}\ \bibnamefont {Santen}}, \bibinfo {author} {\bibfnamefont {B.}~\bibnamefont {Zackay}}, \bibinfo {author} {\bibfnamefont {O.}~\bibnamefont {Guttman}}, \bibinfo {author} {\bibfnamefont {P.}~\bibnamefont {Rekhi}}, \bibinfo {author} {\bibfnamefont {A.}~\bibnamefont {Townsend}}, \bibinfo {author} {\bibfnamefont {A.}~\bibnamefont {Weinstein}}, \ and\ \bibinfo {author} {\bibfnamefont {I.}~\bibnamefont {Wold}},\ }\href {\doibase 10.3847/1538-4357/ad2704} {\bibfield  {journal} {\bibinfo  {journal} {The Astrophysical Journal}\ }\textbf {\bibinfo {volume} {964}},\ \bibinfo {pages} {74} (\bibinfo {year} {2024})}\BibitemShut {NoStop}%
\bibitem [{\citenamefont {{van Velzen}}\ \emph {et~al.}(2021)\citenamefont {{van Velzen}}, \citenamefont {Gezari}, \citenamefont {Hammerstein}, \citenamefont {Roth}, \citenamefont {Frederick}, \citenamefont {Ward}, \citenamefont {Hung}, \citenamefont {Cenko}, \citenamefont {Stein}, \citenamefont {Perley}, \citenamefont {Taggart}, \citenamefont {Foley}, \citenamefont {Sollerman}, \citenamefont {Blagorodnova}, \citenamefont {Andreoni}, \citenamefont {Bellm}, \citenamefont {Brinnel}, \citenamefont {De}, \citenamefont {Dekany}, \citenamefont {Feeney}, \citenamefont {Fremling}, \citenamefont {Giomi}, \citenamefont {Golkhou}, \citenamefont {Graham}, \citenamefont {Ho}, \citenamefont {Kasliwal}, \citenamefont {Kilpatrick}, \citenamefont {Kulkarni}, \citenamefont {Kupfer}, \citenamefont {Laher}, \citenamefont {Mahabal}, \citenamefont {Masci}, \citenamefont {Miller}, \citenamefont {Nordin}, \citenamefont {Riddle}, \citenamefont {Rusholme}, \citenamefont {{van Santen}}, \citenamefont {Sharma}, \citenamefont {Shupe},\
  and\ \citenamefont {Soumagnac}}]{vanvelzenSeventeenTidalDisruption2021}%
  \BibitemOpen
  \bibfield  {author} {\bibinfo {author} {\bibfnamefont {S.}~\bibnamefont {{van Velzen}}}, \bibinfo {author} {\bibfnamefont {S.}~\bibnamefont {Gezari}}, \bibinfo {author} {\bibfnamefont {E.}~\bibnamefont {Hammerstein}}, \bibinfo {author} {\bibfnamefont {N.}~\bibnamefont {Roth}}, \bibinfo {author} {\bibfnamefont {S.}~\bibnamefont {Frederick}}, \bibinfo {author} {\bibfnamefont {C.}~\bibnamefont {Ward}}, \bibinfo {author} {\bibfnamefont {T.}~\bibnamefont {Hung}}, \bibinfo {author} {\bibfnamefont {S.~B.}\ \bibnamefont {Cenko}}, \bibinfo {author} {\bibfnamefont {R.}~\bibnamefont {Stein}}, \bibinfo {author} {\bibfnamefont {D.~A.}\ \bibnamefont {Perley}}, \bibinfo {author} {\bibfnamefont {K.}~\bibnamefont {Taggart}}, \bibinfo {author} {\bibfnamefont {R.~J.}\ \bibnamefont {Foley}}, \bibinfo {author} {\bibfnamefont {J.}~\bibnamefont {Sollerman}}, \bibinfo {author} {\bibfnamefont {N.}~\bibnamefont {Blagorodnova}}, \bibinfo {author} {\bibfnamefont {I.}~\bibnamefont {Andreoni}}, \bibinfo {author} {\bibfnamefont {E.~C.}\
  \bibnamefont {Bellm}}, \bibinfo {author} {\bibfnamefont {V.}~\bibnamefont {Brinnel}}, \bibinfo {author} {\bibfnamefont {K.}~\bibnamefont {De}}, \bibinfo {author} {\bibfnamefont {R.}~\bibnamefont {Dekany}}, \bibinfo {author} {\bibfnamefont {M.}~\bibnamefont {Feeney}}, \bibinfo {author} {\bibfnamefont {C.}~\bibnamefont {Fremling}}, \bibinfo {author} {\bibfnamefont {M.}~\bibnamefont {Giomi}}, \bibinfo {author} {\bibfnamefont {V.~Z.}\ \bibnamefont {Golkhou}}, \bibinfo {author} {\bibfnamefont {M.~J.}\ \bibnamefont {Graham}}, \bibinfo {author} {\bibfnamefont {{\relax Anna}.~Y.~Q.}\ \bibnamefont {Ho}}, \bibinfo {author} {\bibfnamefont {M.~M.}\ \bibnamefont {Kasliwal}}, \bibinfo {author} {\bibfnamefont {C.~D.}\ \bibnamefont {Kilpatrick}}, \bibinfo {author} {\bibfnamefont {S.~R.}\ \bibnamefont {Kulkarni}}, \bibinfo {author} {\bibfnamefont {T.}~\bibnamefont {Kupfer}}, \bibinfo {author} {\bibfnamefont {R.~R.}\ \bibnamefont {Laher}}, \bibinfo {author} {\bibfnamefont {A.}~\bibnamefont {Mahabal}}, \bibinfo {author}
  {\bibfnamefont {F.~J.}\ \bibnamefont {Masci}}, \bibinfo {author} {\bibfnamefont {A.~A.}\ \bibnamefont {Miller}}, \bibinfo {author} {\bibfnamefont {J.}~\bibnamefont {Nordin}}, \bibinfo {author} {\bibfnamefont {R.}~\bibnamefont {Riddle}}, \bibinfo {author} {\bibfnamefont {B.}~\bibnamefont {Rusholme}}, \bibinfo {author} {\bibfnamefont {J.}~\bibnamefont {{van Santen}}}, \bibinfo {author} {\bibfnamefont {Y.}~\bibnamefont {Sharma}}, \bibinfo {author} {\bibfnamefont {D.~L.}\ \bibnamefont {Shupe}}, \ and\ \bibinfo {author} {\bibfnamefont {M.~T.}\ \bibnamefont {Soumagnac}},\ }\href {\doibase 10.3847/1538-4357/abc258} {\bibfield  {journal} {\bibinfo  {journal} {The Astrophysical Journal}\ }\textbf {\bibinfo {volume} {908}},\ \bibinfo {pages} {4} (\bibinfo {year} {2021})}\BibitemShut {NoStop}%
\bibitem [{\citenamefont {Sazonov}\ \emph {et~al.}(2021)\citenamefont {Sazonov}, \citenamefont {Gilfanov}, \citenamefont {Medvedev}, \citenamefont {Yao}, \citenamefont {Khorunzhev}, \citenamefont {Semena}, \citenamefont {Sunyaev}, \citenamefont {Burenin}, \citenamefont {Lyapin}, \citenamefont {Meshcheryakov}, \citenamefont {Uskov}, \citenamefont {Zaznobin}, \citenamefont {Postnov}, \citenamefont {Dodin}, \citenamefont {Belinski}, \citenamefont {Cherepashchuk}, \citenamefont {Eselevich}, \citenamefont {Dodonov}, \citenamefont {Grokhovskaya}, \citenamefont {Kotov}, \citenamefont {Bikmaev}, \citenamefont {Zhuchkov}, \citenamefont {Gumerov}, \citenamefont {{van Velzen}},\ and\ \citenamefont {Kulkarni}}]{sazonovFirstTidalDisruption2021}%
  \BibitemOpen
  \bibfield  {author} {\bibinfo {author} {\bibfnamefont {S.}~\bibnamefont {Sazonov}}, \bibinfo {author} {\bibfnamefont {M.}~\bibnamefont {Gilfanov}}, \bibinfo {author} {\bibfnamefont {P.}~\bibnamefont {Medvedev}}, \bibinfo {author} {\bibfnamefont {Y.}~\bibnamefont {Yao}}, \bibinfo {author} {\bibfnamefont {G.}~\bibnamefont {Khorunzhev}}, \bibinfo {author} {\bibfnamefont {A.}~\bibnamefont {Semena}}, \bibinfo {author} {\bibfnamefont {R.}~\bibnamefont {Sunyaev}}, \bibinfo {author} {\bibfnamefont {R.}~\bibnamefont {Burenin}}, \bibinfo {author} {\bibfnamefont {A.}~\bibnamefont {Lyapin}}, \bibinfo {author} {\bibfnamefont {A.}~\bibnamefont {Meshcheryakov}}, \bibinfo {author} {\bibfnamefont {G.}~\bibnamefont {Uskov}}, \bibinfo {author} {\bibfnamefont {I.}~\bibnamefont {Zaznobin}}, \bibinfo {author} {\bibfnamefont {K.~A.}\ \bibnamefont {Postnov}}, \bibinfo {author} {\bibfnamefont {A.~V.}\ \bibnamefont {Dodin}}, \bibinfo {author} {\bibfnamefont {A.~A.}\ \bibnamefont {Belinski}}, \bibinfo {author} {\bibfnamefont {A.~M.}\
  \bibnamefont {Cherepashchuk}}, \bibinfo {author} {\bibfnamefont {M.}~\bibnamefont {Eselevich}}, \bibinfo {author} {\bibfnamefont {S.~N.}\ \bibnamefont {Dodonov}}, \bibinfo {author} {\bibfnamefont {A.~A.}\ \bibnamefont {Grokhovskaya}}, \bibinfo {author} {\bibfnamefont {S.~S.}\ \bibnamefont {Kotov}}, \bibinfo {author} {\bibfnamefont {I.~F.}\ \bibnamefont {Bikmaev}}, \bibinfo {author} {\bibfnamefont {R.~Y.}\ \bibnamefont {Zhuchkov}}, \bibinfo {author} {\bibfnamefont {R.~I.}\ \bibnamefont {Gumerov}}, \bibinfo {author} {\bibfnamefont {S.}~\bibnamefont {{van Velzen}}}, \ and\ \bibinfo {author} {\bibfnamefont {S.}~\bibnamefont {Kulkarni}},\ }\href {\doibase 10.1093/mnras/stab2843} {\bibfield  {journal} {\bibinfo  {journal} {Monthly Notices of the Royal Astronomical Society}\ }\textbf {\bibinfo {volume} {508}},\ \bibinfo {pages} {3820} (\bibinfo {year} {2021})}\BibitemShut {NoStop}%
\bibitem [{\citenamefont {Lin}\ \emph {et~al.}(2022)\citenamefont {Lin}, \citenamefont {Jiang}, \citenamefont {Kong}, \citenamefont {Huang}, \citenamefont {Lin}, \citenamefont {Zhu},\ and\ \citenamefont {Wang}}]{linLuminosityFunctionTidal2022}%
  \BibitemOpen
  \bibfield  {author} {\bibinfo {author} {\bibfnamefont {Z.}~\bibnamefont {Lin}}, \bibinfo {author} {\bibfnamefont {N.}~\bibnamefont {Jiang}}, \bibinfo {author} {\bibfnamefont {X.}~\bibnamefont {Kong}}, \bibinfo {author} {\bibfnamefont {S.}~\bibnamefont {Huang}}, \bibinfo {author} {\bibfnamefont {Z.}~\bibnamefont {Lin}}, \bibinfo {author} {\bibfnamefont {J.}~\bibnamefont {Zhu}}, \ and\ \bibinfo {author} {\bibfnamefont {Y.}~\bibnamefont {Wang}},\ }\href {\doibase 10.3847/2041-8213/ac9c63} {\bibfield  {journal} {\bibinfo  {journal} {The Astrophysical Journal}\ }\textbf {\bibinfo {volume} {939}},\ \bibinfo {pages} {L33} (\bibinfo {year} {2022})}\BibitemShut {NoStop}%
\bibitem [{\citenamefont {Yao}\ \emph {et~al.}(2023)\citenamefont {Yao}, \citenamefont {Ravi}, \citenamefont {Gezari}, \citenamefont {{van Velzen}}, \citenamefont {Lu}, \citenamefont {Schulze}, \citenamefont {Somalwar}, \citenamefont {Kulkarni}, \citenamefont {Hammerstein}, \citenamefont {Nicholl}, \citenamefont {Graham}, \citenamefont {Perley}, \citenamefont {Cenko}, \citenamefont {Stein}, \citenamefont {Ricarte}, \citenamefont {Chadayammuri}, \citenamefont {Quataert}, \citenamefont {Bellm}, \citenamefont {Bloom}, \citenamefont {Dekany}, \citenamefont {Drake}, \citenamefont {Groom}, \citenamefont {Mahabal}, \citenamefont {Prince}, \citenamefont {Riddle}, \citenamefont {Rusholme}, \citenamefont {Sharma}, \citenamefont {Sollerman},\ and\ \citenamefont {Yan}}]{yaoTidalDisruptionEvent2023}%
  \BibitemOpen
  \bibfield  {author} {\bibinfo {author} {\bibfnamefont {Y.}~\bibnamefont {Yao}}, \bibinfo {author} {\bibfnamefont {V.}~\bibnamefont {Ravi}}, \bibinfo {author} {\bibfnamefont {S.}~\bibnamefont {Gezari}}, \bibinfo {author} {\bibfnamefont {S.}~\bibnamefont {{van Velzen}}}, \bibinfo {author} {\bibfnamefont {W.}~\bibnamefont {Lu}}, \bibinfo {author} {\bibfnamefont {S.}~\bibnamefont {Schulze}}, \bibinfo {author} {\bibfnamefont {J.~J.}\ \bibnamefont {Somalwar}}, \bibinfo {author} {\bibfnamefont {S.~R.}\ \bibnamefont {Kulkarni}}, \bibinfo {author} {\bibfnamefont {E.}~\bibnamefont {Hammerstein}}, \bibinfo {author} {\bibfnamefont {M.}~\bibnamefont {Nicholl}}, \bibinfo {author} {\bibfnamefont {M.~J.}\ \bibnamefont {Graham}}, \bibinfo {author} {\bibfnamefont {D.~A.}\ \bibnamefont {Perley}}, \bibinfo {author} {\bibfnamefont {S.~B.}\ \bibnamefont {Cenko}}, \bibinfo {author} {\bibfnamefont {R.}~\bibnamefont {Stein}}, \bibinfo {author} {\bibfnamefont {A.}~\bibnamefont {Ricarte}}, \bibinfo {author} {\bibfnamefont
  {U.}~\bibnamefont {Chadayammuri}}, \bibinfo {author} {\bibfnamefont {E.}~\bibnamefont {Quataert}}, \bibinfo {author} {\bibfnamefont {E.~C.}\ \bibnamefont {Bellm}}, \bibinfo {author} {\bibfnamefont {J.~S.}\ \bibnamefont {Bloom}}, \bibinfo {author} {\bibfnamefont {R.}~\bibnamefont {Dekany}}, \bibinfo {author} {\bibfnamefont {A.~J.}\ \bibnamefont {Drake}}, \bibinfo {author} {\bibfnamefont {S.~L.}\ \bibnamefont {Groom}}, \bibinfo {author} {\bibfnamefont {A.~A.}\ \bibnamefont {Mahabal}}, \bibinfo {author} {\bibfnamefont {T.~A.}\ \bibnamefont {Prince}}, \bibinfo {author} {\bibfnamefont {R.}~\bibnamefont {Riddle}}, \bibinfo {author} {\bibfnamefont {B.}~\bibnamefont {Rusholme}}, \bibinfo {author} {\bibfnamefont {Y.}~\bibnamefont {Sharma}}, \bibinfo {author} {\bibfnamefont {J.}~\bibnamefont {Sollerman}}, \ and\ \bibinfo {author} {\bibfnamefont {L.}~\bibnamefont {Yan}},\ }\href {\doibase 10.3847/2041-8213/acf216} {\bibfield  {journal} {\bibinfo  {journal} {The Astrophysical Journal}\ }\textbf {\bibinfo {volume}
  {955}},\ \bibinfo {pages} {L6} (\bibinfo {year} {2023})}\BibitemShut {NoStop}%
\bibitem [{\citenamefont {Arcavi}\ \emph {et~al.}(2014)\citenamefont {Arcavi}, \citenamefont {{Gal-Yam}}, \citenamefont {Sullivan}, \citenamefont {Pan}, \citenamefont {Cenko}, \citenamefont {Horesh}, \citenamefont {Ofek}, \citenamefont {Cia}, \citenamefont {Yan}, \citenamefont {Yang}, \citenamefont {Howell}, \citenamefont {Tal}, \citenamefont {Kulkarni}, \citenamefont {Tendulkar}, \citenamefont {Tang}, \citenamefont {Xu}, \citenamefont {Sternberg}, \citenamefont {Cohen}, \citenamefont {Bloom}, \citenamefont {Nugent}, \citenamefont {Kasliwal}, \citenamefont {Perley}, \citenamefont {Quimby}, \citenamefont {Miller}, \citenamefont {Theissen},\ and\ \citenamefont {Laher}}]{arcaviCONTINUUMHeRICHTIDAL2014}%
  \BibitemOpen
  \bibfield  {author} {\bibinfo {author} {\bibfnamefont {I.}~\bibnamefont {Arcavi}}, \bibinfo {author} {\bibfnamefont {A.}~\bibnamefont {{Gal-Yam}}}, \bibinfo {author} {\bibfnamefont {M.}~\bibnamefont {Sullivan}}, \bibinfo {author} {\bibfnamefont {Y.-C.}\ \bibnamefont {Pan}}, \bibinfo {author} {\bibfnamefont {S.~B.}\ \bibnamefont {Cenko}}, \bibinfo {author} {\bibfnamefont {A.}~\bibnamefont {Horesh}}, \bibinfo {author} {\bibfnamefont {E.~O.}\ \bibnamefont {Ofek}}, \bibinfo {author} {\bibfnamefont {A.~D.}\ \bibnamefont {Cia}}, \bibinfo {author} {\bibfnamefont {L.}~\bibnamefont {Yan}}, \bibinfo {author} {\bibfnamefont {C.-W.}\ \bibnamefont {Yang}}, \bibinfo {author} {\bibfnamefont {D.~A.}\ \bibnamefont {Howell}}, \bibinfo {author} {\bibfnamefont {D.}~\bibnamefont {Tal}}, \bibinfo {author} {\bibfnamefont {S.~R.}\ \bibnamefont {Kulkarni}}, \bibinfo {author} {\bibfnamefont {S.~P.}\ \bibnamefont {Tendulkar}}, \bibinfo {author} {\bibfnamefont {S.}~\bibnamefont {Tang}}, \bibinfo {author} {\bibfnamefont
  {D.}~\bibnamefont {Xu}}, \bibinfo {author} {\bibfnamefont {A.}~\bibnamefont {Sternberg}}, \bibinfo {author} {\bibfnamefont {J.~G.}\ \bibnamefont {Cohen}}, \bibinfo {author} {\bibfnamefont {J.~S.}\ \bibnamefont {Bloom}}, \bibinfo {author} {\bibfnamefont {P.~E.}\ \bibnamefont {Nugent}}, \bibinfo {author} {\bibfnamefont {M.~M.}\ \bibnamefont {Kasliwal}}, \bibinfo {author} {\bibfnamefont {D.~A.}\ \bibnamefont {Perley}}, \bibinfo {author} {\bibfnamefont {R.~M.}\ \bibnamefont {Quimby}}, \bibinfo {author} {\bibfnamefont {A.~A.}\ \bibnamefont {Miller}}, \bibinfo {author} {\bibfnamefont {C.~A.}\ \bibnamefont {Theissen}}, \ and\ \bibinfo {author} {\bibfnamefont {R.~R.}\ \bibnamefont {Laher}},\ }\href {\doibase 10.1088/0004-637X/793/1/38} {\bibfield  {journal} {\bibinfo  {journal} {The Astrophysical Journal}\ }\textbf {\bibinfo {volume} {793}},\ \bibinfo {pages} {38} (\bibinfo {year} {2014})}\BibitemShut {NoStop}%
\bibitem [{\citenamefont {French}\ \emph {et~al.}(2016)\citenamefont {French}, \citenamefont {Arcavi},\ and\ \citenamefont {Zabludoff}}]{frenchTidalDisruptionEvents2016}%
  \BibitemOpen
  \bibfield  {author} {\bibinfo {author} {\bibfnamefont {K.~D.}\ \bibnamefont {French}}, \bibinfo {author} {\bibfnamefont {I.}~\bibnamefont {Arcavi}}, \ and\ \bibinfo {author} {\bibfnamefont {A.}~\bibnamefont {Zabludoff}},\ }\href {\doibase 10.3847/2041-8205/818/1/L21} {\bibfield  {journal} {\bibinfo  {journal} {The Astrophysical Journal}\ }\textbf {\bibinfo {volume} {818}},\ \bibinfo {pages} {L21} (\bibinfo {year} {2016})}\BibitemShut {NoStop}%
\bibitem [{\citenamefont {{Law-Smith}}\ \emph {et~al.}(2017)\citenamefont {{Law-Smith}}, \citenamefont {{Ramirez-Ruiz}}, \citenamefont {Ellison},\ and\ \citenamefont {Foley}}]{law-smithTidalDisruptionEvent2017}%
  \BibitemOpen
  \bibfield  {author} {\bibinfo {author} {\bibfnamefont {J.}~\bibnamefont {{Law-Smith}}}, \bibinfo {author} {\bibfnamefont {E.}~\bibnamefont {{Ramirez-Ruiz}}}, \bibinfo {author} {\bibfnamefont {S.~L.}\ \bibnamefont {Ellison}}, \ and\ \bibinfo {author} {\bibfnamefont {R.~J.}\ \bibnamefont {Foley}},\ }\href {\doibase 10.3847/1538-4357/aa94c7} {\bibfield  {journal} {\bibinfo  {journal} {The Astrophysical Journal}\ }\textbf {\bibinfo {volume} {850}},\ \bibinfo {pages} {22} (\bibinfo {year} {2017})}\BibitemShut {NoStop}%
\bibitem [{\citenamefont {Wang}\ and\ \citenamefont {Merritt}(2004)}]{wangRevisedRatesStellar2004}%
  \BibitemOpen
  \bibfield  {author} {\bibinfo {author} {\bibfnamefont {J.}~\bibnamefont {Wang}}\ and\ \bibinfo {author} {\bibfnamefont {D.}~\bibnamefont {Merritt}},\ }\href {\doibase 10.1086/379767} {\bibfield  {journal} {\bibinfo  {journal} {The Astrophysical Journal}\ }\textbf {\bibinfo {volume} {600}},\ \bibinfo {pages} {149} (\bibinfo {year} {2004})},\ \Eprint {http://arxiv.org/abs/astro-ph/0305493} {arXiv:astro-ph/0305493} \BibitemShut {NoStop}%
\bibitem [{\citenamefont {Stone}\ and\ \citenamefont {Metzger}(2016)}]{stoneRatesStellarTidal2016}%
  \BibitemOpen
  \bibfield  {author} {\bibinfo {author} {\bibfnamefont {N.~C.}\ \bibnamefont {Stone}}\ and\ \bibinfo {author} {\bibfnamefont {B.~D.}\ \bibnamefont {Metzger}},\ }\href {\doibase 10.1093/mnras/stv2281} {\bibfield  {journal} {\bibinfo  {journal} {Monthly Notices of the Royal Astronomical Society}\ }\textbf {\bibinfo {volume} {455}},\ \bibinfo {pages} {859} (\bibinfo {year} {2016})},\ \Eprint {http://arxiv.org/abs/1410.7772} {arXiv:1410.7772} \BibitemShut {NoStop}%
\bibitem [{\citenamefont {Teboul}\ \emph {et~al.}(2022)\citenamefont {Teboul}, \citenamefont {Stone},\ and\ \citenamefont {Ostriker}}]{teboulLossConeShielding2022}%
  \BibitemOpen
  \bibfield  {author} {\bibinfo {author} {\bibfnamefont {O.}~\bibnamefont {Teboul}}, \bibinfo {author} {\bibfnamefont {N.~C.}\ \bibnamefont {Stone}}, \ and\ \bibinfo {author} {\bibfnamefont {J.~P.}\ \bibnamefont {Ostriker}},\ }\href@noop {} {\enquote {\bibinfo {title} {Loss {{Cone Shielding}}},}\ } (\bibinfo {year} {2022}),\ \Eprint {http://arxiv.org/abs/2211.05858} {arXiv:2211.05858 [astro-ph]} \BibitemShut {NoStop}%
\bibitem [{\citenamefont {Broggi}\ \emph {et~al.}(2024)\citenamefont {Broggi}, \citenamefont {Stone}, \citenamefont {Ryu}, \citenamefont {Bortolas}, \citenamefont {Dotti}, \citenamefont {Bonetti},\ and\ \citenamefont {Sesana}}]{broggiRepeatingPartialDisruptions2024}%
  \BibitemOpen
  \bibfield  {author} {\bibinfo {author} {\bibfnamefont {L.}~\bibnamefont {Broggi}}, \bibinfo {author} {\bibfnamefont {N.~C.}\ \bibnamefont {Stone}}, \bibinfo {author} {\bibfnamefont {T.}~\bibnamefont {Ryu}}, \bibinfo {author} {\bibfnamefont {E.}~\bibnamefont {Bortolas}}, \bibinfo {author} {\bibfnamefont {M.}~\bibnamefont {Dotti}}, \bibinfo {author} {\bibfnamefont {M.}~\bibnamefont {Bonetti}}, \ and\ \bibinfo {author} {\bibfnamefont {A.}~\bibnamefont {Sesana}},\ }\href {\doibase 10.33232/001c.120086} {\bibfield  {journal} {\bibinfo  {journal} {The Open Journal of Astrophysics}\ }\textbf {\bibinfo {volume} {7}},\ \bibinfo {pages} {48} (\bibinfo {year} {2024})}\BibitemShut {NoStop}%
\bibitem [{\citenamefont {Stone}\ \emph {et~al.}(2018)\citenamefont {Stone}, \citenamefont {Generozov}, \citenamefont {Vasiliev},\ and\ \citenamefont {Metzger}}]{stoneDelayTimeDistribution2018}%
  \BibitemOpen
  \bibfield  {author} {\bibinfo {author} {\bibfnamefont {N.~C.}\ \bibnamefont {Stone}}, \bibinfo {author} {\bibfnamefont {A.}~\bibnamefont {Generozov}}, \bibinfo {author} {\bibfnamefont {E.}~\bibnamefont {Vasiliev}}, \ and\ \bibinfo {author} {\bibfnamefont {B.~D.}\ \bibnamefont {Metzger}},\ }\href {\doibase 10.1093/mnras/sty2045} {\bibfield  {journal} {\bibinfo  {journal} {Monthly Notices of the Royal Astronomical Society}\ }\textbf {\bibinfo {volume} {480}},\ \bibinfo {pages} {5060} (\bibinfo {year} {2018})}\BibitemShut {NoStop}%
\bibitem [{\citenamefont {Bortolas}(2022)}]{bortolasTidalDisruptionEvents2022}%
  \BibitemOpen
  \bibfield  {author} {\bibinfo {author} {\bibfnamefont {E.}~\bibnamefont {Bortolas}},\ }\href {\doibase 10.1093/mnras/stac262} {\  (\bibinfo {year} {2022}),\ 10.1093/mnras/stac262}\BibitemShut {NoStop}%
\bibitem [{\citenamefont {Wang}\ \emph {et~al.}(2024)\citenamefont {Wang}, \citenamefont {Ma}, \citenamefont {Wu},\ and\ \citenamefont {Jiang}}]{wangExplanationOverrepresentationTidal2024}%
  \BibitemOpen
  \bibfield  {author} {\bibinfo {author} {\bibfnamefont {M.}~\bibnamefont {Wang}}, \bibinfo {author} {\bibfnamefont {Y.}~\bibnamefont {Ma}}, \bibinfo {author} {\bibfnamefont {Q.}~\bibnamefont {Wu}}, \ and\ \bibinfo {author} {\bibfnamefont {N.}~\bibnamefont {Jiang}},\ }\href {\doibase 10.3847/1538-4357/ad0bfb} {\bibfield  {journal} {\bibinfo  {journal} {The Astrophysical Journal}\ }\textbf {\bibinfo {volume} {960}},\ \bibinfo {pages} {69} (\bibinfo {year} {2024})}\BibitemShut {NoStop}%
\bibitem [{\citenamefont {Babak}\ \emph {et~al.}(2017)\citenamefont {Babak}, \citenamefont {Gair}, \citenamefont {Sesana}, \citenamefont {Barausse}, \citenamefont {Sopuerta}, \citenamefont {Berry}, \citenamefont {Berti}, \citenamefont {{Amaro-Seoane}}, \citenamefont {Petiteau},\ and\ \citenamefont {Klein}}]{babakScienceSpacebasedInterferometer2017}%
  \BibitemOpen
  \bibfield  {author} {\bibinfo {author} {\bibfnamefont {S.}~\bibnamefont {Babak}}, \bibinfo {author} {\bibfnamefont {J.}~\bibnamefont {Gair}}, \bibinfo {author} {\bibfnamefont {A.}~\bibnamefont {Sesana}}, \bibinfo {author} {\bibfnamefont {E.}~\bibnamefont {Barausse}}, \bibinfo {author} {\bibfnamefont {C.~F.}\ \bibnamefont {Sopuerta}}, \bibinfo {author} {\bibfnamefont {C.~P.~L.}\ \bibnamefont {Berry}}, \bibinfo {author} {\bibfnamefont {E.}~\bibnamefont {Berti}}, \bibinfo {author} {\bibfnamefont {P.}~\bibnamefont {{Amaro-Seoane}}}, \bibinfo {author} {\bibfnamefont {A.}~\bibnamefont {Petiteau}}, \ and\ \bibinfo {author} {\bibfnamefont {A.}~\bibnamefont {Klein}},\ }\href {\doibase 10.1103/PhysRevD.95.103012} {\bibfield  {journal} {\bibinfo  {journal} {Physical Review D}\ }\textbf {\bibinfo {volume} {95}},\ \bibinfo {pages} {103012} (\bibinfo {year} {2017})}\BibitemShut {NoStop}%
\bibitem [{\citenamefont {{Amaro-Seoane}}(2018)}]{amaro-seoaneRelativisticDynamicsExtreme2018a}%
  \BibitemOpen
  \bibfield  {author} {\bibinfo {author} {\bibfnamefont {P.}~\bibnamefont {{Amaro-Seoane}}},\ }\href {\doibase 10.1007/s41114-018-0013-8} {\bibfield  {journal} {\bibinfo  {journal} {Living Reviews in Relativity}\ }\textbf {\bibinfo {volume} {21}},\ \bibinfo {pages} {4} (\bibinfo {year} {2018})}\BibitemShut {NoStop}%
\bibitem [{\citenamefont {Broggi}\ \emph {et~al.}(2022)\citenamefont {Broggi}, \citenamefont {Bortolas}, \citenamefont {Bonetti}, \citenamefont {Sesana},\ and\ \citenamefont {Dotti}}]{broggiExtremeMassRatio2022}%
  \BibitemOpen
  \bibfield  {author} {\bibinfo {author} {\bibfnamefont {L.}~\bibnamefont {Broggi}}, \bibinfo {author} {\bibfnamefont {E.}~\bibnamefont {Bortolas}}, \bibinfo {author} {\bibfnamefont {M.}~\bibnamefont {Bonetti}}, \bibinfo {author} {\bibfnamefont {A.}~\bibnamefont {Sesana}}, \ and\ \bibinfo {author} {\bibfnamefont {M.}~\bibnamefont {Dotti}},\ }\href {\doibase 10.1093/mnras/stac1453} {\bibfield  {journal} {\bibinfo  {journal} {Monthly Notices of the Royal Astronomical Society}\ }\textbf {\bibinfo {volume} {514}},\ \bibinfo {pages} {3270} (\bibinfo {year} {2022})}\BibitemShut {NoStop}%
\bibitem [{\citenamefont {Rom}\ \emph {et~al.}(2024)\citenamefont {Rom}, \citenamefont {Linial}, \citenamefont {Kaur},\ and\ \citenamefont {Sari}}]{romDynamicsSupermassiveBlack2024}%
  \BibitemOpen
  \bibfield  {author} {\bibinfo {author} {\bibfnamefont {B.}~\bibnamefont {Rom}}, \bibinfo {author} {\bibfnamefont {I.}~\bibnamefont {Linial}}, \bibinfo {author} {\bibfnamefont {K.}~\bibnamefont {Kaur}}, \ and\ \bibinfo {author} {\bibfnamefont {R.}~\bibnamefont {Sari}},\ }\href {\doibase 10.48550/arXiv.2406.19443} {\enquote {\bibinfo {title} {Dynamics around supermassive black holes: {{Extreme}} mass-ratio inspirals as gravitational-wave sources},}\ } (\bibinfo {year} {2024})\BibitemShut {NoStop}%
\bibitem [{\citenamefont {Colpi}\ \emph {et~al.}(2024)\citenamefont {Colpi}, \citenamefont {Danzmann}, \citenamefont {Hewitson}, \citenamefont {{Holley-Bockelmann}}, \citenamefont {Jetzer}, \citenamefont {Nelemans}, \citenamefont {Petiteau}, \citenamefont {Shoemaker}, \citenamefont {Sopuerta}, \citenamefont {Stebbins}, \citenamefont {Tanvir}, \citenamefont {Ward}, \citenamefont {Weber}, \citenamefont {Thorpe}, \citenamefont {Daurskikh}, \citenamefont {Deep}, \citenamefont {Fern{\'a}ndez~N{\'u}{\~n}ez}, \citenamefont {Garc{\'i}a~Marirrodriga}, \citenamefont {Gehler}, \citenamefont {Halain}, \citenamefont {Jennrich}, \citenamefont {Lammers}, \citenamefont {Larra{\~n}aga}, \citenamefont {Lieser}, \citenamefont {L{\"u}tzgendorf}, \citenamefont {Martens}, \citenamefont {Mondin}, \citenamefont {Piris~Ni{\~n}o}, \citenamefont {{Amaro-Seoane}}, \citenamefont {Arca~Sedda}, \citenamefont {Auclair}, \citenamefont {Babak}, \citenamefont {Baghi}, \citenamefont {Baibhav}, \citenamefont {Baker}, \citenamefont {Bayle},
  \citenamefont {Berry}, \citenamefont {Berti}, \citenamefont {Boileau}, \citenamefont {Bonetti}, \citenamefont {Brito}, \citenamefont {Buscicchio}, \citenamefont {Calcagni}, \citenamefont {Capelo}, \citenamefont {Caprini}, \citenamefont {Caputo}, \citenamefont {Castelli}, \citenamefont {Chen}, \citenamefont {Chen}, \citenamefont {Chua}, \citenamefont {Davies}, \citenamefont {Derdzinski}, \citenamefont {Domcke}, \citenamefont {Doneva}, \citenamefont {Dvorkin}, \citenamefont {Mar{\'i}a~Ezquiaga}, \citenamefont {Gair}, \citenamefont {Haiman}, \citenamefont {Harry}, \citenamefont {Hartwig}, \citenamefont {Hees}, \citenamefont {Heffernan}, \citenamefont {Husa}, \citenamefont {{Izquierdo-Villalba}}, \citenamefont {Karnesis}, \citenamefont {Klein}, \citenamefont {Korol}, \citenamefont {Korsakova}, \citenamefont {Kupfer}, \citenamefont {Laghi}, \citenamefont {Lamberts}, \citenamefont {Larson}, \citenamefont {Le~Jeune}, \citenamefont {Lewicki}, \citenamefont {Littenberg}, \citenamefont {Madge}, \citenamefont
  {Mangiagli}, \citenamefont {Marsat}, \citenamefont {Vilchez}, \citenamefont {Maselli}, \citenamefont {Mathews}, \citenamefont {{van de Meent}}, \citenamefont {Muratore}, \citenamefont {Nardini}, \citenamefont {Pani}, \citenamefont {Peloso}, \citenamefont {Pieroni}, \citenamefont {Pound}, \citenamefont {{Quelquejay-Leclere}}, \citenamefont {Ricciardone}, \citenamefont {Rossi}, \citenamefont {Sartirana}, \citenamefont {Savalle}, \citenamefont {Sberna}, \citenamefont {Sesana}, \citenamefont {Shoemaker}, \citenamefont {Slutsky}, \citenamefont {Sotiriou}, \citenamefont {Speri}, \citenamefont {Staab}, \citenamefont {Steer}, \citenamefont {Tamanini}, \citenamefont {Tasinato}, \citenamefont {Torrado}, \citenamefont {{Torres-Orjuela}}, \citenamefont {Toubiana}, \citenamefont {Vallisneri}, \citenamefont {Vecchio}, \citenamefont {Volonteri}, \citenamefont {Yagi},\ and\ \citenamefont {Zwick}}]{colpiLISADefinitionStudy2024}%
  \BibitemOpen
  \bibfield  {author} {\bibinfo {author} {\bibfnamefont {M.}~\bibnamefont {Colpi}}, \bibinfo {author} {\bibfnamefont {K.}~\bibnamefont {Danzmann}}, \bibinfo {author} {\bibfnamefont {M.}~\bibnamefont {Hewitson}}, \bibinfo {author} {\bibfnamefont {K.}~\bibnamefont {{Holley-Bockelmann}}}, \bibinfo {author} {\bibfnamefont {P.}~\bibnamefont {Jetzer}}, \bibinfo {author} {\bibfnamefont {G.}~\bibnamefont {Nelemans}}, \bibinfo {author} {\bibfnamefont {A.}~\bibnamefont {Petiteau}}, \bibinfo {author} {\bibfnamefont {D.}~\bibnamefont {Shoemaker}}, \bibinfo {author} {\bibfnamefont {C.}~\bibnamefont {Sopuerta}}, \bibinfo {author} {\bibfnamefont {R.}~\bibnamefont {Stebbins}}, \bibinfo {author} {\bibfnamefont {N.}~\bibnamefont {Tanvir}}, \bibinfo {author} {\bibfnamefont {H.}~\bibnamefont {Ward}}, \bibinfo {author} {\bibfnamefont {W.~J.}\ \bibnamefont {Weber}}, \bibinfo {author} {\bibfnamefont {I.}~\bibnamefont {Thorpe}}, \bibinfo {author} {\bibfnamefont {A.}~\bibnamefont {Daurskikh}}, \bibinfo {author} {\bibfnamefont
  {A.}~\bibnamefont {Deep}}, \bibinfo {author} {\bibfnamefont {I.}~\bibnamefont {Fern{\'a}ndez~N{\'u}{\~n}ez}}, \bibinfo {author} {\bibfnamefont {C.}~\bibnamefont {Garc{\'i}a~Marirrodriga}}, \bibinfo {author} {\bibfnamefont {M.}~\bibnamefont {Gehler}}, \bibinfo {author} {\bibfnamefont {J.-P.}\ \bibnamefont {Halain}}, \bibinfo {author} {\bibfnamefont {O.}~\bibnamefont {Jennrich}}, \bibinfo {author} {\bibfnamefont {U.}~\bibnamefont {Lammers}}, \bibinfo {author} {\bibfnamefont {J.}~\bibnamefont {Larra{\~n}aga}}, \bibinfo {author} {\bibfnamefont {M.}~\bibnamefont {Lieser}}, \bibinfo {author} {\bibfnamefont {N.}~\bibnamefont {L{\"u}tzgendorf}}, \bibinfo {author} {\bibfnamefont {W.}~\bibnamefont {Martens}}, \bibinfo {author} {\bibfnamefont {L.}~\bibnamefont {Mondin}}, \bibinfo {author} {\bibfnamefont {A.}~\bibnamefont {Piris~Ni{\~n}o}}, \bibinfo {author} {\bibfnamefont {P.}~\bibnamefont {{Amaro-Seoane}}}, \bibinfo {author} {\bibfnamefont {M.}~\bibnamefont {Arca~Sedda}}, \bibinfo {author} {\bibfnamefont
  {P.}~\bibnamefont {Auclair}}, \bibinfo {author} {\bibfnamefont {S.}~\bibnamefont {Babak}}, \bibinfo {author} {\bibfnamefont {Q.}~\bibnamefont {Baghi}}, \bibinfo {author} {\bibfnamefont {V.}~\bibnamefont {Baibhav}}, \bibinfo {author} {\bibfnamefont {T.}~\bibnamefont {Baker}}, \bibinfo {author} {\bibfnamefont {J.-B.}\ \bibnamefont {Bayle}}, \bibinfo {author} {\bibfnamefont {C.}~\bibnamefont {Berry}}, \bibinfo {author} {\bibfnamefont {E.}~\bibnamefont {Berti}}, \bibinfo {author} {\bibfnamefont {G.}~\bibnamefont {Boileau}}, \bibinfo {author} {\bibfnamefont {M.}~\bibnamefont {Bonetti}}, \bibinfo {author} {\bibfnamefont {R.}~\bibnamefont {Brito}}, \bibinfo {author} {\bibfnamefont {R.}~\bibnamefont {Buscicchio}}, \bibinfo {author} {\bibfnamefont {G.}~\bibnamefont {Calcagni}}, \bibinfo {author} {\bibfnamefont {P.~R.}\ \bibnamefont {Capelo}}, \bibinfo {author} {\bibfnamefont {C.}~\bibnamefont {Caprini}}, \bibinfo {author} {\bibfnamefont {A.}~\bibnamefont {Caputo}}, \bibinfo {author} {\bibfnamefont {E.}~\bibnamefont
  {Castelli}}, \bibinfo {author} {\bibfnamefont {H.-Y.}\ \bibnamefont {Chen}}, \bibinfo {author} {\bibfnamefont {X.}~\bibnamefont {Chen}}, \bibinfo {author} {\bibfnamefont {A.}~\bibnamefont {Chua}}, \bibinfo {author} {\bibfnamefont {G.}~\bibnamefont {Davies}}, \bibinfo {author} {\bibfnamefont {A.}~\bibnamefont {Derdzinski}}, \bibinfo {author} {\bibfnamefont {V.~F.}\ \bibnamefont {Domcke}}, \bibinfo {author} {\bibfnamefont {D.}~\bibnamefont {Doneva}}, \bibinfo {author} {\bibfnamefont {I.}~\bibnamefont {Dvorkin}}, \bibinfo {author} {\bibfnamefont {J.}~\bibnamefont {Mar{\'i}a~Ezquiaga}}, \bibinfo {author} {\bibfnamefont {J.}~\bibnamefont {Gair}}, \bibinfo {author} {\bibfnamefont {Z.}~\bibnamefont {Haiman}}, \bibinfo {author} {\bibfnamefont {I.}~\bibnamefont {Harry}}, \bibinfo {author} {\bibfnamefont {O.}~\bibnamefont {Hartwig}}, \bibinfo {author} {\bibfnamefont {A.}~\bibnamefont {Hees}}, \bibinfo {author} {\bibfnamefont {A.}~\bibnamefont {Heffernan}}, \bibinfo {author} {\bibfnamefont {S.}~\bibnamefont {Husa}},
  \bibinfo {author} {\bibfnamefont {D.}~\bibnamefont {{Izquierdo-Villalba}}}, \bibinfo {author} {\bibfnamefont {N.}~\bibnamefont {Karnesis}}, \bibinfo {author} {\bibfnamefont {A.}~\bibnamefont {Klein}}, \bibinfo {author} {\bibfnamefont {V.}~\bibnamefont {Korol}}, \bibinfo {author} {\bibfnamefont {N.}~\bibnamefont {Korsakova}}, \bibinfo {author} {\bibfnamefont {T.}~\bibnamefont {Kupfer}}, \bibinfo {author} {\bibfnamefont {D.}~\bibnamefont {Laghi}}, \bibinfo {author} {\bibfnamefont {A.}~\bibnamefont {Lamberts}}, \bibinfo {author} {\bibfnamefont {S.}~\bibnamefont {Larson}}, \bibinfo {author} {\bibfnamefont {M.}~\bibnamefont {Le~Jeune}}, \bibinfo {author} {\bibfnamefont {M.}~\bibnamefont {Lewicki}}, \bibinfo {author} {\bibfnamefont {T.}~\bibnamefont {Littenberg}}, \bibinfo {author} {\bibfnamefont {E.}~\bibnamefont {Madge}}, \bibinfo {author} {\bibfnamefont {A.}~\bibnamefont {Mangiagli}}, \bibinfo {author} {\bibfnamefont {S.}~\bibnamefont {Marsat}}, \bibinfo {author} {\bibfnamefont {I.~M.}\ \bibnamefont
  {Vilchez}}, \bibinfo {author} {\bibfnamefont {A.}~\bibnamefont {Maselli}}, \bibinfo {author} {\bibfnamefont {J.}~\bibnamefont {Mathews}}, \bibinfo {author} {\bibfnamefont {M.}~\bibnamefont {{van de Meent}}}, \bibinfo {author} {\bibfnamefont {M.}~\bibnamefont {Muratore}}, \bibinfo {author} {\bibfnamefont {G.}~\bibnamefont {Nardini}}, \bibinfo {author} {\bibfnamefont {P.}~\bibnamefont {Pani}}, \bibinfo {author} {\bibfnamefont {M.}~\bibnamefont {Peloso}}, \bibinfo {author} {\bibfnamefont {M.}~\bibnamefont {Pieroni}}, \bibinfo {author} {\bibfnamefont {A.}~\bibnamefont {Pound}}, \bibinfo {author} {\bibfnamefont {H.}~\bibnamefont {{Quelquejay-Leclere}}}, \bibinfo {author} {\bibfnamefont {A.}~\bibnamefont {Ricciardone}}, \bibinfo {author} {\bibfnamefont {E.~M.}\ \bibnamefont {Rossi}}, \bibinfo {author} {\bibfnamefont {A.}~\bibnamefont {Sartirana}}, \bibinfo {author} {\bibfnamefont {E.}~\bibnamefont {Savalle}}, \bibinfo {author} {\bibfnamefont {L.}~\bibnamefont {Sberna}}, \bibinfo {author} {\bibfnamefont
  {A.}~\bibnamefont {Sesana}}, \bibinfo {author} {\bibfnamefont {D.}~\bibnamefont {Shoemaker}}, \bibinfo {author} {\bibfnamefont {J.}~\bibnamefont {Slutsky}}, \bibinfo {author} {\bibfnamefont {T.}~\bibnamefont {Sotiriou}}, \bibinfo {author} {\bibfnamefont {L.}~\bibnamefont {Speri}}, \bibinfo {author} {\bibfnamefont {M.}~\bibnamefont {Staab}}, \bibinfo {author} {\bibfnamefont {D.}~\bibnamefont {Steer}}, \bibinfo {author} {\bibfnamefont {N.}~\bibnamefont {Tamanini}}, \bibinfo {author} {\bibfnamefont {G.}~\bibnamefont {Tasinato}}, \bibinfo {author} {\bibfnamefont {J.}~\bibnamefont {Torrado}}, \bibinfo {author} {\bibfnamefont {A.}~\bibnamefont {{Torres-Orjuela}}}, \bibinfo {author} {\bibfnamefont {A.}~\bibnamefont {Toubiana}}, \bibinfo {author} {\bibfnamefont {M.}~\bibnamefont {Vallisneri}}, \bibinfo {author} {\bibfnamefont {A.}~\bibnamefont {Vecchio}}, \bibinfo {author} {\bibfnamefont {M.}~\bibnamefont {Volonteri}}, \bibinfo {author} {\bibfnamefont {K.}~\bibnamefont {Yagi}}, \ and\ \bibinfo {author}
  {\bibfnamefont {L.}~\bibnamefont {Zwick}},\ }\href {\doibase 10.48550/arXiv.2402.07571} {\enquote {\bibinfo {title} {{{LISA Definition Study Report}}},}\ } (\bibinfo {year} {2024})\BibitemShut {NoStop}%
\bibitem [{\citenamefont {Binney}\ and\ \citenamefont {Tremaine}(2008)}]{binneyGalacticDynamicsSecond2008}%
  \BibitemOpen
  \bibfield  {author} {\bibinfo {author} {\bibfnamefont {J.}~\bibnamefont {Binney}}\ and\ \bibinfo {author} {\bibfnamefont {S.}~\bibnamefont {Tremaine}},\ }\href@noop {} {\emph {\bibinfo {title} {Galactic {{Dynamics}}: {{Second Edition}}}}}\ (\bibinfo {year} {2008})\BibitemShut {NoStop}%
\bibitem [{\citenamefont {Merritt}(2013)}]{merrittDynamicsEvolutionGalactic2013}%
  \BibitemOpen
  \bibfield  {author} {\bibinfo {author} {\bibfnamefont {D.}~\bibnamefont {Merritt}},\ }\href@noop {} {\emph {\bibinfo {title} {Dynamics and {{Evolution}} of {{Galactic Nuclei}}}}}\ (\bibinfo  {publisher} {Princeton University Press},\ \bibinfo {year} {2013})\BibitemShut {NoStop}%
\bibitem [{\citenamefont {Cohn}\ and\ \citenamefont {Kulsrud}(1978)}]{cohnStellarDistributionBlack1978}%
  \BibitemOpen
  \bibfield  {author} {\bibinfo {author} {\bibfnamefont {H.}~\bibnamefont {Cohn}}\ and\ \bibinfo {author} {\bibfnamefont {R.~M.}\ \bibnamefont {Kulsrud}},\ }\href {\doibase 10.1086/156685} {\bibfield  {journal} {\bibinfo  {journal} {The Astrophysical Journal}\ }\textbf {\bibinfo {volume} {226}},\ \bibinfo {pages} {1087} (\bibinfo {year} {1978})}\BibitemShut {NoStop}%
\bibitem [{\citenamefont {Mancieri}\ \emph {et~al.}(2024)\citenamefont {Mancieri}, \citenamefont {Broggi}, \citenamefont {Bonetti},\ and\ \citenamefont {Sesana}}]{mancieriHangingCliffEMRI2024}%
  \BibitemOpen
  \bibfield  {author} {\bibinfo {author} {\bibfnamefont {D.}~\bibnamefont {Mancieri}}, \bibinfo {author} {\bibfnamefont {L.}~\bibnamefont {Broggi}}, \bibinfo {author} {\bibfnamefont {M.}~\bibnamefont {Bonetti}}, \ and\ \bibinfo {author} {\bibfnamefont {A.}~\bibnamefont {Sesana}},\ }\href {\doibase 10.48550/arXiv.2409.09122} {\enquote {\bibinfo {title} {Hanging on the cliff: {{EMRI}} formation with local two-body relaxation and post-{{Newtonian}} dynamics},}\ } (\bibinfo {year} {2024}),\ \Eprint {http://arxiv.org/abs/2409.09122} {arXiv:2409.09122 [astro-ph, physics:gr-qc]} \BibitemShut {NoStop}%
\bibitem [{\citenamefont {Pan}\ and\ \citenamefont {Yang}(2021)}]{panFormationRateExtreme2021}%
  \BibitemOpen
  \bibfield  {author} {\bibinfo {author} {\bibfnamefont {Z.}~\bibnamefont {Pan}}\ and\ \bibinfo {author} {\bibfnamefont {H.}~\bibnamefont {Yang}},\ }\href@noop {} {\bibfield  {journal} {\bibinfo  {journal} {arXiv:2101.09146 [astro-ph, physics:gr-qc]}\ } (\bibinfo {year} {2021})},\ \Eprint {http://arxiv.org/abs/2101.09146} {arXiv:2101.09146 [astro-ph, physics:gr-qc]} \BibitemShut {NoStop}%
\bibitem [{\citenamefont {Wang}\ \emph {et~al.}(2023)\citenamefont {Wang}, \citenamefont {Ma},\ and\ \citenamefont {Wu}}]{wangAccretionmodifiedStellarmassBlack2023}%
  \BibitemOpen
  \bibfield  {author} {\bibinfo {author} {\bibfnamefont {M.}~\bibnamefont {Wang}}, \bibinfo {author} {\bibfnamefont {Y.}~\bibnamefont {Ma}}, \ and\ \bibinfo {author} {\bibfnamefont {Q.}~\bibnamefont {Wu}},\ }\href {\doibase 10.1093/mnras/stad422} {\bibfield  {journal} {\bibinfo  {journal} {Monthly Notices of the Royal Astronomical Society}\ }\textbf {\bibinfo {volume} {520}},\ \bibinfo {pages} {4502} (\bibinfo {year} {2023})}\BibitemShut {NoStop}%
\bibitem [{\citenamefont {Sari}\ and\ \citenamefont {Fragione}(2019)}]{sariTidalDisruptionEvents2019}%
  \BibitemOpen
  \bibfield  {author} {\bibinfo {author} {\bibfnamefont {R.}~\bibnamefont {Sari}}\ and\ \bibinfo {author} {\bibfnamefont {G.}~\bibnamefont {Fragione}},\ }\href {\doibase 10.3847/1538-4357/ab43df} {\bibfield  {journal} {\bibinfo  {journal} {The Astrophysical Journal}\ }\textbf {\bibinfo {volume} {885}},\ \bibinfo {pages} {24} (\bibinfo {year} {2019})}\BibitemShut {NoStop}%
\bibitem [{\citenamefont {Qunbar}\ and\ \citenamefont {Stone}(2023)}]{qunbarEnhancedExtremeMass2023}%
  \BibitemOpen
  \bibfield  {author} {\bibinfo {author} {\bibfnamefont {I.}~\bibnamefont {Qunbar}}\ and\ \bibinfo {author} {\bibfnamefont {N.~C.}\ \bibnamefont {Stone}},\ }\href@noop {} {\enquote {\bibinfo {title} {Enhanced {{Extreme Mass Ratio Inspiral Rates}} into {{Intermediate Mass Black Holes}}},}\ } (\bibinfo {year} {2023}),\ \Eprint {http://arxiv.org/abs/2304.13062} {arXiv:2304.13062 [astro-ph, physics:gr-qc]} \BibitemShut {NoStop}%
\bibitem [{\citenamefont {Kolomgorov}\ \emph {et~al.}(1937)\citenamefont {Kolomgorov}, \citenamefont {Petrovskii},\ and\ \citenamefont {Piskunov}}]{kolomgorovStudyDiffusionEquation1937}%
  \BibitemOpen
  \bibfield  {author} {\bibinfo {author} {\bibfnamefont {A.~N.}\ \bibnamefont {Kolomgorov}}, \bibinfo {author} {\bibfnamefont {I.~G.}\ \bibnamefont {Petrovskii}}, \ and\ \bibinfo {author} {\bibfnamefont {N.~S.}\ \bibnamefont {Piskunov}},\ }\href@noop {} {\bibfield  {journal} {\bibinfo  {journal} {Mosk. Gos. Univ. Ser. A Mat. Mekh}\ }\textbf {\bibinfo {volume} {1}},\ \bibinfo {pages} {26} (\bibinfo {year} {1937})}\BibitemShut {NoStop}%
\bibitem [{\citenamefont {Pfister}\ \emph {et~al.}(2022)\citenamefont {Pfister}, \citenamefont {Toscani}, \citenamefont {Wong}, \citenamefont {Dai}, \citenamefont {Lodato},\ and\ \citenamefont {Rossi}}]{pfisterObservableGravitationalWaves2022}%
  \BibitemOpen
  \bibfield  {author} {\bibinfo {author} {\bibfnamefont {H.}~\bibnamefont {Pfister}}, \bibinfo {author} {\bibfnamefont {M.}~\bibnamefont {Toscani}}, \bibinfo {author} {\bibfnamefont {T.~H.~T.}\ \bibnamefont {Wong}}, \bibinfo {author} {\bibfnamefont {J.~L.}\ \bibnamefont {Dai}}, \bibinfo {author} {\bibfnamefont {G.}~\bibnamefont {Lodato}}, \ and\ \bibinfo {author} {\bibfnamefont {E.~M.}\ \bibnamefont {Rossi}},\ }\href {\doibase 10.1093/mnras/stab3387} {\bibfield  {journal} {\bibinfo  {journal} {Monthly Notices of the Royal Astronomical Society}\ }\textbf {\bibinfo {volume} {510}},\ \bibinfo {pages} {2025} (\bibinfo {year} {2022})}\BibitemShut {NoStop}%
\bibitem [{\citenamefont {Chang}\ \emph {et~al.}(2024)\citenamefont {Chang}, \citenamefont {Dai}, \citenamefont {Pfister}, \citenamefont {Chowdhury},\ and\ \citenamefont {Natarajan}}]{changRatesStellarTidal2024}%
  \BibitemOpen
  \bibfield  {author} {\bibinfo {author} {\bibfnamefont {J.~N.~Y.}\ \bibnamefont {Chang}}, \bibinfo {author} {\bibfnamefont {L.}~\bibnamefont {Dai}}, \bibinfo {author} {\bibfnamefont {H.}~\bibnamefont {Pfister}}, \bibinfo {author} {\bibfnamefont {R.~K.}\ \bibnamefont {Chowdhury}}, \ and\ \bibinfo {author} {\bibfnamefont {P.}~\bibnamefont {Natarajan}},\ }\href {\doibase 10.48550/arXiv.2407.09339} {\enquote {\bibinfo {title} {Rates of {{Stellar Tidal Disruption Events Around Intermediate-Mass Black Holes}}},}\ } (\bibinfo {year} {2024})\BibitemShut {NoStop}%
\bibitem [{\citenamefont {Bortolas}\ \emph {et~al.}(2023)\citenamefont {Bortolas}, \citenamefont {Ryu}, \citenamefont {Broggi},\ and\ \citenamefont {Sesana}}]{bortolasPartialStellarTidal2023}%
  \BibitemOpen
  \bibfield  {author} {\bibinfo {author} {\bibfnamefont {E.}~\bibnamefont {Bortolas}}, \bibinfo {author} {\bibfnamefont {T.}~\bibnamefont {Ryu}}, \bibinfo {author} {\bibfnamefont {L.}~\bibnamefont {Broggi}}, \ and\ \bibinfo {author} {\bibfnamefont {A.}~\bibnamefont {Sesana}},\ }\href@noop {} {\enquote {\bibinfo {title} {Partial stellar tidal disruption events and their rates},}\ } (\bibinfo {year} {2023}),\ \Eprint {http://arxiv.org/abs/2303.03408} {arXiv:2303.03408 [astro-ph]} \BibitemShut {NoStop}%
\bibitem [{\citenamefont {Vasiliev}\ and\ \citenamefont {Merritt}(2013)}]{vasilievLossconeProblemAxisymmetric2013}%
  \BibitemOpen
  \bibfield  {author} {\bibinfo {author} {\bibfnamefont {E.}~\bibnamefont {Vasiliev}}\ and\ \bibinfo {author} {\bibfnamefont {D.}~\bibnamefont {Merritt}},\ }\href {\doibase 10.1088/0004-637X/774/1/87} {\bibfield  {journal} {\bibinfo  {journal} {The Astrophysical Journal}\ }\textbf {\bibinfo {volume} {774}},\ \bibinfo {pages} {87} (\bibinfo {year} {2013})}\BibitemShut {NoStop}%
\bibitem [{\citenamefont {Mazzolari}\ \emph {et~al.}(2022)\citenamefont {Mazzolari}, \citenamefont {Bonetti}, \citenamefont {Sesana}, \citenamefont {Colombo}, \citenamefont {Dotti}, \citenamefont {Lodato},\ and\ \citenamefont {{Izquierdo-Villalba}}}]{mazzolariExtremeMassRatio2022}%
  \BibitemOpen
  \bibfield  {author} {\bibinfo {author} {\bibfnamefont {G.}~\bibnamefont {Mazzolari}}, \bibinfo {author} {\bibfnamefont {M.}~\bibnamefont {Bonetti}}, \bibinfo {author} {\bibfnamefont {A.}~\bibnamefont {Sesana}}, \bibinfo {author} {\bibfnamefont {R.~M.}\ \bibnamefont {Colombo}}, \bibinfo {author} {\bibfnamefont {M.}~\bibnamefont {Dotti}}, \bibinfo {author} {\bibfnamefont {G.}~\bibnamefont {Lodato}}, \ and\ \bibinfo {author} {\bibfnamefont {D.}~\bibnamefont {{Izquierdo-Villalba}}},\ }\href {\doibase 10.1093/mnras/stac2255} {\bibfield  {journal} {\bibinfo  {journal} {Monthly Notices of the Royal Astronomical Society}\ }\textbf {\bibinfo {volume} {516}},\ \bibinfo {pages} {1959} (\bibinfo {year} {2022})}\BibitemShut {NoStop}%
\bibitem [{\citenamefont {Liu}\ \emph {et~al.}(2023)\citenamefont {Liu}, \citenamefont {Mockler}, \citenamefont {{Ramirez-Ruiz}}, \citenamefont {Yarza}, \citenamefont {{Law-Smith}}, \citenamefont {Naoz}, \citenamefont {Melchor},\ and\ \citenamefont {Rose}}]{liuTidalDisruptionEvents2023}%
  \BibitemOpen
  \bibfield  {author} {\bibinfo {author} {\bibfnamefont {C.}~\bibnamefont {Liu}}, \bibinfo {author} {\bibfnamefont {B.}~\bibnamefont {Mockler}}, \bibinfo {author} {\bibfnamefont {E.}~\bibnamefont {{Ramirez-Ruiz}}}, \bibinfo {author} {\bibfnamefont {R.}~\bibnamefont {Yarza}}, \bibinfo {author} {\bibfnamefont {J.~A.~P.}\ \bibnamefont {{Law-Smith}}}, \bibinfo {author} {\bibfnamefont {S.}~\bibnamefont {Naoz}}, \bibinfo {author} {\bibfnamefont {D.}~\bibnamefont {Melchor}}, \ and\ \bibinfo {author} {\bibfnamefont {S.}~\bibnamefont {Rose}},\ }\href {\doibase 10.3847/1538-4357/acafe1} {\bibfield  {journal} {\bibinfo  {journal} {The Astrophysical Journal}\ }\textbf {\bibinfo {volume} {944}},\ \bibinfo {pages} {184} (\bibinfo {year} {2023})}\BibitemShut {NoStop}%
\bibitem [{\citenamefont {Kaur}\ and\ \citenamefont {Perets}(2024)}]{kaurExtrememassRatioInspirals2024}%
  \BibitemOpen
  \bibfield  {author} {\bibinfo {author} {\bibfnamefont {K.}~\bibnamefont {Kaur}}\ and\ \bibinfo {author} {\bibfnamefont {H.}~\bibnamefont {Perets}},\ }\href {\doibase 10.48550/arXiv.2409.10618} {\enquote {\bibinfo {title} {Extreme-mass ratio inspirals in strong segregation regime -- to inspiral or to get ejected?}}\ } (\bibinfo {year} {2024})\BibitemShut {NoStop}%
\bibitem [{\citenamefont {Kaur}\ \emph {et~al.}(2024)\citenamefont {Kaur}, \citenamefont {Rom},\ and\ \citenamefont {Sari}}]{kaurSemiAnalyticalFokkerPlanck2024}%
  \BibitemOpen
  \bibfield  {author} {\bibinfo {author} {\bibfnamefont {K.}~\bibnamefont {Kaur}}, \bibinfo {author} {\bibfnamefont {B.}~\bibnamefont {Rom}}, \ and\ \bibinfo {author} {\bibfnamefont {R.}~\bibnamefont {Sari}},\ }\href {\doibase 10.48550/arXiv.2406.07627} {\enquote {\bibinfo {title} {Semi-{{Analytical Fokker Planck Models}} for {{Nuclear Star Clusters}}},}\ } (\bibinfo {year} {2024})\BibitemShut {NoStop}%
\bibitem [{\citenamefont {Bahcall}\ and\ \citenamefont {Wolf}(1977)}]{bahcallStarDistributionMassive1977}%
  \BibitemOpen
  \bibfield  {author} {\bibinfo {author} {\bibfnamefont {J.~N.}\ \bibnamefont {Bahcall}}\ and\ \bibinfo {author} {\bibfnamefont {R.~A.}\ \bibnamefont {Wolf}},\ }\href {\doibase 10.1086/155534} {\bibfield  {journal} {\bibinfo  {journal} {The Astrophysical Journal}\ }\textbf {\bibinfo {volume} {216}},\ \bibinfo {pages} {883} (\bibinfo {year} {1977})}\BibitemShut {NoStop}%
\bibitem [{\citenamefont {Linial}\ and\ \citenamefont {Sari}(2022)}]{linialStellarDistributionsSupermassive2022}%
  \BibitemOpen
  \bibfield  {author} {\bibinfo {author} {\bibfnamefont {I.}~\bibnamefont {Linial}}\ and\ \bibinfo {author} {\bibfnamefont {R.}~\bibnamefont {Sari}},\ }\href {\doibase 10.3847/1538-4357/ac9bfd} {\bibfield  {journal} {\bibinfo  {journal} {The Astrophysical Journal}\ }\textbf {\bibinfo {volume} {940}},\ \bibinfo {pages} {101} (\bibinfo {year} {2022})}\BibitemShut {NoStop}%
\bibitem [{\citenamefont {Rom}\ \emph {et~al.}(2023)\citenamefont {Rom}, \citenamefont {Linial},\ and\ \citenamefont {Sari}}]{romEnergyFluxParticle2023}%
  \BibitemOpen
  \bibfield  {author} {\bibinfo {author} {\bibfnamefont {B.}~\bibnamefont {Rom}}, \bibinfo {author} {\bibfnamefont {I.}~\bibnamefont {Linial}}, \ and\ \bibinfo {author} {\bibfnamefont {R.}~\bibnamefont {Sari}},\ }\href {\doibase 10.3847/1538-4357/acd54f} {\bibfield  {journal} {\bibinfo  {journal} {The Astrophysical Journal}\ }\textbf {\bibinfo {volume} {951}},\ \bibinfo {pages} {14} (\bibinfo {year} {2023})}\BibitemShut {NoStop}%
\bibitem [{\citenamefont {Vasiliev}(2017)}]{vasilievNewFokkerPlanckApproach2017}%
  \BibitemOpen
  \bibfield  {author} {\bibinfo {author} {\bibfnamefont {E.}~\bibnamefont {Vasiliev}},\ }\href {\doibase 10.3847/1538-4357/aa8cc8} {\bibfield  {journal} {\bibinfo  {journal} {arXiv:1709.04467 [astro-ph]}\ } (\bibinfo {year} {2017}),\ 10.3847/1538-4357/aa8cc8},\ \Eprint {http://arxiv.org/abs/1709.04467} {arXiv:1709.04467 [astro-ph]} \BibitemShut {NoStop}%
\bibitem [{\citenamefont {{Holley-Bockelmann}}\ and\ \citenamefont {Sigurdsson}(2006)}]{holley-bockelmannFullLossCone2006}%
  \BibitemOpen
  \bibfield  {author} {\bibinfo {author} {\bibfnamefont {K.}~\bibnamefont {{Holley-Bockelmann}}}\ and\ \bibinfo {author} {\bibfnamefont {S.}~\bibnamefont {Sigurdsson}},\ }\href {\doibase 10.48550/arXiv.astro-ph/0601520} {\enquote {\bibinfo {title} {A {{Full Loss Cone For Triaxial Galaxies}}},}\ } (\bibinfo {year} {2006}),\ \Eprint {http://arxiv.org/abs/astro-ph/0601520} {arXiv:astro-ph/0601520} \BibitemShut {NoStop}%
\bibitem [{\citenamefont {Lezhnin}\ and\ \citenamefont {Vasiliev}(2016)}]{lezhninTIDALDISRUPTIONRATES2016}%
  \BibitemOpen
  \bibfield  {author} {\bibinfo {author} {\bibfnamefont {K.}~\bibnamefont {Lezhnin}}\ and\ \bibinfo {author} {\bibfnamefont {E.}~\bibnamefont {Vasiliev}},\ }\href {\doibase 10.3847/0004-637X/831/1/84} {\bibfield  {journal} {\bibinfo  {journal} {The Astrophysical Journal}\ }\textbf {\bibinfo {volume} {831}},\ \bibinfo {pages} {84} (\bibinfo {year} {2016})}\BibitemShut {NoStop}%
\bibitem [{\citenamefont {Rozner}\ and\ \citenamefont {{Ramirez-Ruiz}}(2025)}]{roznerStellarDistributionsSupermassive2025}%
  \BibitemOpen
  \bibfield  {author} {\bibinfo {author} {\bibfnamefont {M.}~\bibnamefont {Rozner}}\ and\ \bibinfo {author} {\bibfnamefont {E.}~\bibnamefont {{Ramirez-Ruiz}}},\ }\href {\doibase 10.3847/2041-8213/adeca7} {\bibfield  {journal} {\bibinfo  {journal} {The Astrophysical Journal Letters}\ }\textbf {\bibinfo {volume} {988}},\ \bibinfo {pages} {L21} (\bibinfo {year} {2025})}\BibitemShut {NoStop}%
\bibitem [{\citenamefont {Vasiliev}(2014)}]{vasilievRatesCaptureStars2014}%
  \BibitemOpen
  \bibfield  {author} {\bibinfo {author} {\bibfnamefont {E.}~\bibnamefont {Vasiliev}},\ }\href {\doibase 10.1088/0264-9381/31/24/244002} {\bibfield  {journal} {\bibinfo  {journal} {Classical and Quantum Gravity}\ }\textbf {\bibinfo {volume} {31}},\ \bibinfo {pages} {244002} (\bibinfo {year} {2014})}\BibitemShut {NoStop}%
\bibitem [{\citenamefont {Bezanson}\ \emph {et~al.}(2017)\citenamefont {Bezanson}, \citenamefont {Edelman}, \citenamefont {Karpinski},\ and\ \citenamefont {Shah}}]{bezansonJuliaFreshApproach2017}%
  \BibitemOpen
  \bibfield  {author} {\bibinfo {author} {\bibfnamefont {J.}~\bibnamefont {Bezanson}}, \bibinfo {author} {\bibfnamefont {A.}~\bibnamefont {Edelman}}, \bibinfo {author} {\bibfnamefont {S.}~\bibnamefont {Karpinski}}, \ and\ \bibinfo {author} {\bibfnamefont {V.~B.}\ \bibnamefont {Shah}},\ }\href {\doibase 10.1137/141000671} {\bibfield  {journal} {\bibinfo  {journal} {SIAM Review}\ }\textbf {\bibinfo {volume} {59}},\ \bibinfo {pages} {65} (\bibinfo {year} {2017})}\BibitemShut {NoStop}%
\bibitem [{\citenamefont {Danisch}\ and\ \citenamefont {Krumbiegel}(2021)}]{danischMakieJlFlexible2021}%
  \BibitemOpen
  \bibfield  {author} {\bibinfo {author} {\bibfnamefont {S.}~\bibnamefont {Danisch}}\ and\ \bibinfo {author} {\bibfnamefont {J.}~\bibnamefont {Krumbiegel}},\ }\href {\doibase 10.21105/joss.03349} {\bibfield  {journal} {\bibinfo  {journal} {Journal of Open Source Software}\ }\textbf {\bibinfo {volume} {6}},\ \bibinfo {pages} {3349} (\bibinfo {year} {2021})}\BibitemShut {NoStop}%
\bibitem [{\citenamefont {Kittisopikul}\ and\ \citenamefont {Holy}(2023)}]{kittisopikulJuliaMathInterpolationsJl2023}%
  \BibitemOpen
  \bibfield  {author} {\bibinfo {author} {\bibfnamefont {M.}~\bibnamefont {Kittisopikul}}\ and\ \bibinfo {author} {\bibfnamefont {T.~E.}\ \bibnamefont {Holy}},\ }\href@noop {} {\enquote {\bibinfo {title} {{{JuliaMath}}/{{Interpolations}}.jl},}\ } (\bibinfo {year} {2023})\BibitemShut {NoStop}%
\bibitem [{\citenamefont {Milosavljevi{\'c}}\ and\ \citenamefont {Merritt}(2003)}]{milosavljevicLongTermEvolutionMassive2003}%
  \BibitemOpen
  \bibfield  {author} {\bibinfo {author} {\bibfnamefont {M.}~\bibnamefont {Milosavljevi{\'c}}}\ and\ \bibinfo {author} {\bibfnamefont {D.}~\bibnamefont {Merritt}},\ }\href {\doibase 10.1086/378086} {\bibfield  {journal} {\bibinfo  {journal} {The Astrophysical Journal}\ }\textbf {\bibinfo {volume} {596}},\ \bibinfo {pages} {860} (\bibinfo {year} {2003})}\BibitemShut {NoStop}%
\bibitem [{\citenamefont {Broggi}(2024)}]{broggiDynamicsTidalDisruptions}%
  \BibitemOpen
  \bibfield  {author} {\bibinfo {author} {\bibfnamefont {L.}~\bibnamefont {Broggi}},\ }\href {https://hdl.handle.net/10281/459238} {\emph {\bibinfo {title} {Dynamics of Tidal Disruptions and Extreme Mass Ratio Inspirals in Galactic Nuclei}}}\ (\bibinfo {year} {2024})\BibitemShut {NoStop}%
\end{thebibliography}%

\appendix

\section{Relaxed profile from the local Fokker-Planck equation}
\label{ap:ap}

\citetalias{cohnStellarDistributionBlack1978} start from the local Fokker-Planck equation Eq.~\eqref{eq:local} in terms of $(E, \R, r)$, considering separately $v_r >0$ ($f^\mathrm{out}$) and $v_r < 0$ ($f^\mathrm{in}$). They show that the collisional term for $\R \to 0$ reduces to
\begin{equation}\label{eq:local_inside}
\begin{split}
    \Gamma_\mathrm{coll} &\simeq -\frac{\partial}{\partial \R} F_R= \frac{\partial}{\partial \R} d_{\R\R} \, \frac{\partial}{\partial \R} f^\mathrm{in/out} \\ d_{\R\R} &= \R\, {d}(E,r)
\end{split}
\end{equation}
which is the form we used in Eq.~\eqref{eq:FP_inside}. Then, they change variable\footnote{The partial derivative $\partial_\R$ in Eq.~\eqref{eq:local_inside} is at fixed $r$, and the term should transform when changing variable to $\tau^\mathrm{CK}$ since $r(\R, \tau^\mathrm{CK}) \neq r(\R + \mathrm{d}\R, \tau^\mathrm{CK})$. We maintain $\tau^\mathrm{CK}$ instead of changing variable to $\tau$ to keep the same level of approximation.} from $\R$ to $y = \R/\R_\LC$ and from $r$ to $\tau^\mathrm{CK}\in[0, 1]$, an analogous to $\tau$ from Eq~\eqref{eq:tau},
\begin{equation}\label{eq:tau_CK}
    \tau^\mathrm{CK}(r) = \frac{1}{\D \, P}\times\begin{cases}
        \int_{r_p}^{r} \frac{\mathrm{d}r}{v_r}\, d &\quad v_r \geq 0\\
        \mathcal{D} P + \int_{r_p}^{r} \frac{\mathrm{d}r}{v_r}\, d  &\quad v_r < 0
    \end{cases}
\end{equation}
to treat $f^\mathrm{in/out}$ together; the steady-state equation reduces to
\begin{equation}
    \frac{\partial}{\partial \tau^\mathrm{CK}} f(E; y, \tau^\mathrm{CK}) = q\,\frac{\partial}{\partial y}\,y\frac{\partial}{\partial y} f (E; y, \tau^\mathrm{CK})\, .
\end{equation}
The equation is solved in the domain $0 \leq \tau^\mathrm{CK} < 1$ and $0 \leq y \leq y_\mathrm{max}$. The upper bound to $y$ is such that $y_\mathrm{max} > 1$ (\textit{i.e.} $\R_\mathrm{max} > \R_\LC$), and the boundary conditions from the problem are:

\medskip
\begin{tabular}{rl}
    a. & $\qquad f(E; y, 0) = f(E;y,1)$ for $y>y_\mathrm{max}$\\
    b. & $\qquad f(E; y, 0) = 0$ for $y<y_\mathrm{max}$\\
    c. & $\qquad f(E; y_\mathrm{max}, \tau^\mathrm{CK}) = f_\LC$\\
    d. & $\qquad F_R(E; 0, \tau^\mathrm{CK}) = 0$
\end{tabular}
\medskip

\begin{figure}\centering
    \includegraphics[width=\linewidth]{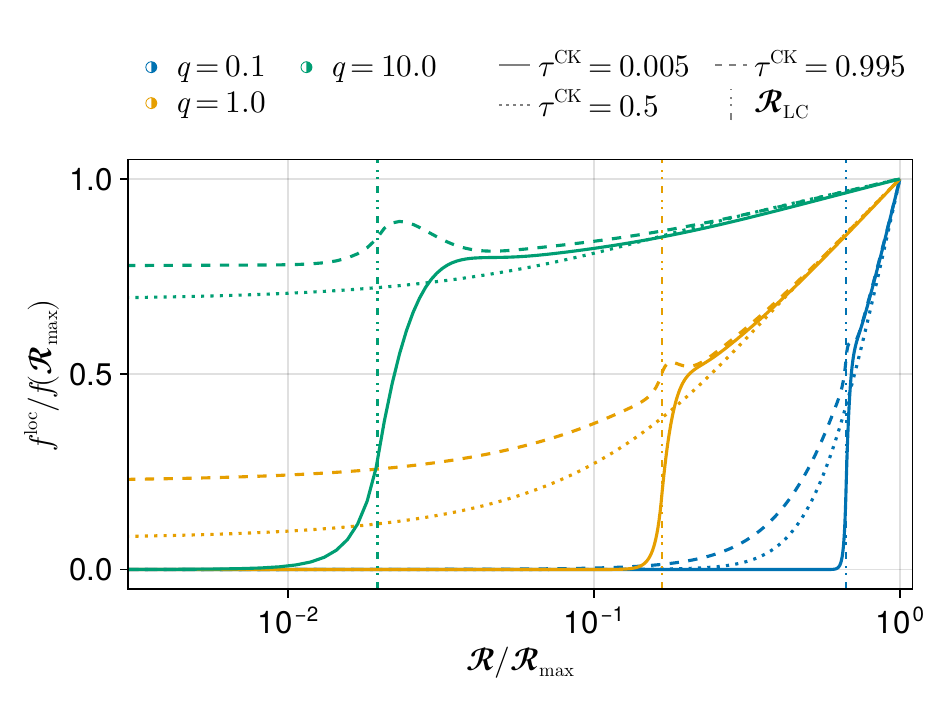}
    \caption{\label{fig:ymax_test}Numerical solution inside the loss cone for $\tau^\mathrm{CK} = 0.005$ (solid), $\tau^\mathrm{CK} = 0.5$ (dot) and $\tau^\mathrm{CK}=0.995$ (dash) for $q=0.1$ (blue), $q=1.0$ (orange) and $q=10.0$ (green). The vertical lines mark the loss cone value $\R_\mathrm{LC}$. As $q$ increases, we increase $\R_\mathrm{max}$ according to Eq.~\eqref{eq:ymax}, as close to $\R_\mathrm{max}$ the distribution becomes uniform in $\tau$.}
\end{figure}

As argued by \citetalias{cohnStellarDistributionBlack1978}, the value of $y_\mathrm{max}$ should be much larger than $1.0$, as the function will asymptotically be uniform in $\tau^\mathrm{CK}$ to match the solution of the OAFPE. As $y$ approaches $1^+$, $f$ will start showing an explicit dependence on $\tau^\mathrm{CK}$.
We solve the equation by representing $\tau^\mathrm{CK}$ and $y$ on a grid with $N_\tau$ points in $\tau^\mathrm{CK}$, and $N_y$ points in the interval $0<y<y_\mathrm{max}$. We write the steady-state equation using finite difference as
\begin{equation}
    \mathfrak{A} \cdot \mathfrak{f} = \mathfrak{b}
\end{equation}
where $\mathfrak{b}$ is set by the initial/boundary conditions.
We consider $q = \{0.1, 1.0, 10.0\}$, and setting $N_\tau = 200$ we find that the solution becomes insensitive to $\R_\mathrm{max}$ for
\begin{equation}\label{eq:ymax}
    y_\mathrm{max} = \R_\mathrm{max} / \R_\LC \gtrsim 1 + 5\, q \,.
\end{equation}
This formula has a simple interpretation: $q\, \R_\LC$ is the amplitude $\sigma$ of the ``kicks'' received over an orbital period, and less than one star in a million is expected to end up at $5\,\sigma = 5 \, q\, \R_\LC$ from $\R_\LC$. In this work, we set $y_\mathrm{max}$ according to Eq.~\eqref{eq:ymax}, and set $N_\tau=200$ and $N_y = 800$.  This resolution ensures that, for any value of $q$ explored, inside and outside the loss cone there are at least $20$ grid points. In Fig.~\ref{fig:ymax_test} we show the solution for $q=\{0.1, 1.0, 10.0\}$; our choice of $\R_\mathrm{max}$ ensures that the DF is uniform in $\tau$ at large $\R$, and approaches a logarithm.

\paragraph{Analytic solution for $\R_\mathrm{max} = \R_\LC$.}
Assuming $y_\mathrm{max} = y_\LC$, the solution can be expressed as an infinite series \footnote{The solution was originally derived by M. Milosavljevic following a related derivation of \citet{milosavljevicLongTermEvolutionMassive2003}, later presented in L.Strubbe's PhD thesis (2011), and published in \citet{merrittDynamicsEvolutionGalactic2013}. See Chap.3 of \citet{broggiDynamicsTidalDisruptions} for a detailed derivation based on the separation of variables.}.
\begin{equation}
    \frac{f^\mathrm{M}}{f_\LC} =  1 - 2\, \sum_{k=1}^{\infty} \frac{\exp\{-\tau^\mathrm{CK}\,q\,\alpha^2_k/4\}}{\alpha_k \, J_1(\alpha_k) } \, J_0\left(\alpha_k \, \sqrt{\frac{\R}{\R_\LC} }\right)
\end{equation}
where $J_n$ is the Bessel function of the first kind of order $n$, and $\alpha_k$ is the $k$-th zero of $J_0$.

The average distribution predicted is
\begin{equation}
    \begin{split}\label{eq:avg_local_FP}
        \frac{f^\mathrm{M}_\mathrm{avg}}{f_\LC}=1 - 8\sum_{k=1}^{\infty} \frac{1 - \exp\{-q\,\alpha^2_k/4\}}{\alpha_k^3 \, J_1(\alpha_k) } \, J_0\left(\alpha_k \, \sqrt{\frac{\R}{\R_\LC} }\right)
    \end{split}
\end{equation}
and the distribution at disruption is
\begin{equation}\label{eq:cap_local_FP}
    f^\mathrm{M}_\mathrm{cap} = f^\mathrm{M}(E,\R,\tau^\mathrm{CK} = 1).
\end{equation}
The latter gives the total flux entering the loss cone boundary as in Eq.~\eqref{eq:FLC_SS}
\begin{equation}\label{eq:Flux_local_FP}
    \begin{split}
        \F_\LC =  4\,\pi^2\,L_c^2 \, \R_\LC\,f_\LC\left[1-\sum_{k=1}^{\infty} \frac{4}{\alpha_k^2}\exp\left(-\frac{q\,\alpha_k^2}{4}\right)\right] \, .
    \end{split}
\end{equation}
In this work we use the first 200 terms of the series, which ensure convergence for $\tau^\mathrm{CK} \gtrsim 0.002$.

\begin{figure}
\includegraphics[width=\linewidth]{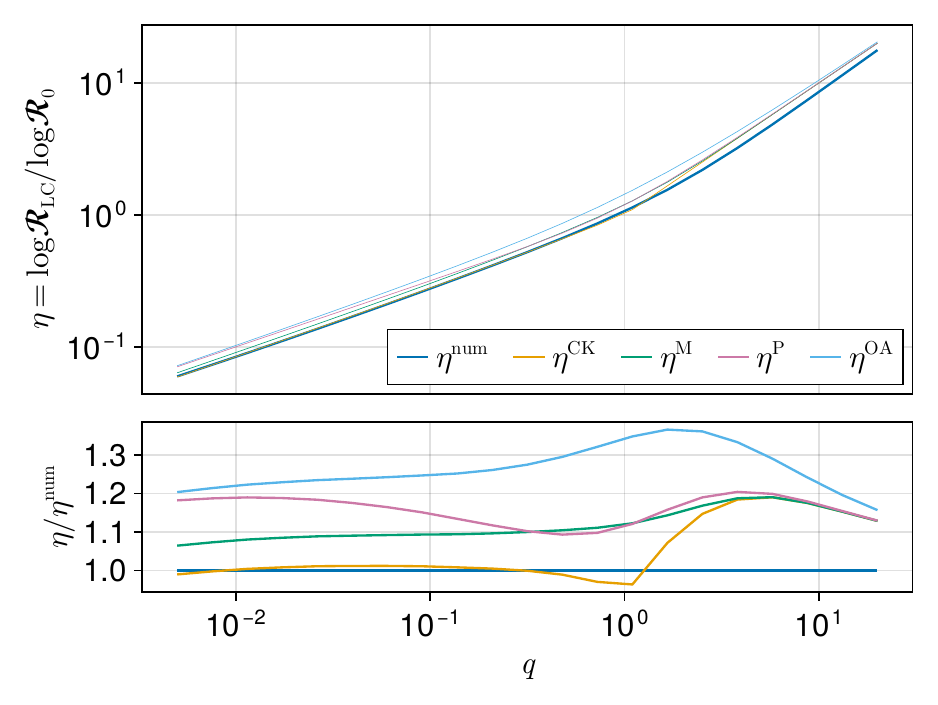}
    \caption{\label{fig:comparison_eta} Value of $\eta = \log \R_\LC / \R_0$ from the numerical solution ($\eta^\mathrm{num}$, blue), the fit by \citetalias{cohnStellarDistributionBlack1978} ($\eta^\mathrm{CK}$, Eq.~\eqref{eq:eta_CK}, orange), the prescription by \citet{merrittDynamicsEvolutionGalactic2013}  ($\eta^\mathrm{M}$, Eq.~\eqref{eq:eta_M}, green), the simplified expression ($\eta^\mathrm{P}$, Eq.~\eqref{eq:eta_P}, pink) and the orbit-averaged estimate  ($\eta^\mathrm{OA}$, Eq.~\eqref{eq:eta}, light blue).}
\end{figure}

\section{Immediate applications of the orbit-averaged approach}\label{ap:montecarlo}
\paragraph{Orbit-averaged Monte Carlo.} Monte Carlo codes to solve the OAFPE in a period-by-period fashion can accurately account for loss cone captures only when  $t_\mathrm{rlx}>P$. In fact, a star is immediately removed by the system as soon as $\R < \R_\LC$ \citep[see ][sec. 4.2, for an example where the period-by-period approach fails]{mancieriHangingCliffEMRI2024}.  Eq.~\eqref{eq:sink_term} suggests a prescription to account for the full loss cone regime. As $t_\rlx$ sets the time for relaxation to change orbital angular momentum, a star is captured if it is scattered at a point of the orbit such that
\begin{equation}
    P-t_\mathrm{rlx} < \tau \leq P,
\end{equation}
and, in the orbit-averaged approximation, this occurs with a probability
\begin{equation}
    p = \min{\left(\frac{t_\mathrm{rlx}}{P}, 1\right)} \, .
\end{equation}
By introducing $q$ of Eq.~\eqref{eq:q}, $t_\mathrm{rlx} \simeq P \, \R/(q \R_\LC )$, and therefore
\begin{equation}
    p = \min{\left(\frac{\R}{q(E)\,\R_\LC(E)}, 1\right)} \, .
\end{equation}
In the full loss cone regime this probability is smaller than one at the loss cone, and will go to zero for radial orbits. In the empty loss cone regime, most plunges occur at $\R\simeq \R_\LC$, and $p = 1$ as expected.

\paragraph{Boundary conditions}
The solution inside the loss cone relates the value of the DF at the loss cone $f_\LC$ and the flux of disruptions $\F_\LC$, classically setting a boundary condition for the OAFPE.
Specifically, one matches the solution inside the loss cone to the steady-state form of $f$ outside of it
\begin{equation}
\begin{split}\label{eq:log_outside}
        f &\sim f_\LC \, \frac{\log \R/\R_0}{\log \R_\LC / \R_0} \qquad \R \to \R_\mathrm{max}^+
\end{split}
\end{equation}
A direct comparison with Eq.~\eqref{eq:Flux} (superscript OA) and Eq.~\eqref{eq:Flux_local_FP} (superscript $\mathrm{M}$) gives the following estimates for $\eta \equiv \log \R_\LC / \R_0$
\begin{align}
    \eta^\mathrm{OA} &= \sqrt{q} \; \frac{I_0(2/\sqrt{q})}{I_1(2/\sqrt{q})} \label{eq:eta} \, ,\\
    \eta^\mathrm{M} &= q \, \left[1-\sum_{k=1}^{\infty} \frac{4}{\alpha_k^2}\exp\left(-\frac{q\,\alpha_k^2}{4}\right)\right]^{-1} \, .\label{eq:eta_M}
\end{align}

\citetalias{cohnStellarDistributionBlack1978} and \citet{merrittDynamicsEvolutionGalactic2013} (superscript P)\footnote{The 1D Fokker-Planck code \textsc{PhaseFlow} uses $\eta^\mathrm{P}$ \citep{vasilievNewFokkerPlanckApproach2017}} report some approximants
\begin{align}
    \eta^\mathrm{CK} &= \begin{cases}
        q &\qquad q\geq 1\\
        0.186\,q + 0.824\, \sqrt{q} &\qquad q<1
    \end{cases}\label{eq:eta_CK}\\
    \eta^\mathrm{P} &= \sqrt[4]{q^2+q^4} \label{eq:eta_P} \, .
\end{align}
We compare the various estimates in Fig.\ref{fig:comparison_eta}. Analytic prescriptions overestimate $\eta$, except for $\eta^\mathrm{CK}$ at $q\lesssim1$. The orbit-averaged estimate has the largest deviations, contained within $35\%$. Since $q$ and $\eta$ range orders of magnitude in a stellar system \citep{merrittDynamicsEvolutionGalactic2013}, all estimates will be equivalent for the solution of the OAFPE on a grid.

\end{document}